\documentclass[fleqn,usenatbib]{mnras}

\usepackage{newtxtext,newtxmath}

\usepackage[T1]{fontenc}

\DeclareRobustCommand{\VAN}[3]{#2}
\let\VANthebibliography\thebibliography
\def\thebibliography{\DeclareRobustCommand{\VAN}[3]{##3}\VANthebibliography}

\usepackage{graphicx}	
\usepackage{amsmath}	
\usepackage[normalem]{ulem}
\usepackage{color}
\usepackage{gensymb}
\usepackage{subcaption}
\usepackage{makecell}
\usepackage{hyperref}
\usepackage{url}
\usepackage{orcidlink}

\newcommand{\prob}[4]{\mathcal{P}^{#1}_{\text{#2}}(#3|N_{\text{det}}^{#4})}
\newcommand{\ftprob}[4]{\tilde{\mathcal{P}}^{#1}_{\text{#2}}(#3|N_{\text{det}}{#4})}

\newcommand{\ndet}[1]{N_{\text{det}}=#1}

\title[Comparison of Cross-Correlation Methods for LIM]{Comparison of Cross-Correlation Methods for Line Intensity Mapping}

\author[S. H. Kramer et al.]{
Samuel H. Kramer \orcidlink{0009-0009-6060-108X},$^{1}$\thanks{e-mail: shkramer@wisc.edu}
Patrick C. Breysse \orcidlink{0000-0001-8382-5275},$^{2}$
Anthony R. Pullen \orcidlink{0000-0002-2091-8738},$^{3,4}$
Faizah K. Siddique \orcidlink{0009-0005-8009-1583},$^{1}$
\newauthor{ Eric R. Switzer,$^{5}$
Peter T. Timbie \orcidlink{0000-0003-0325-1633},$^{1}$ and Dongwoo Chung \orcidlink{0000-0003-2618-6504}$^{6}$}
\\
$^{1}$Department of Physics, University of Wisconsin-Madison, 1150 University Avenue, Madison, WI 53706, United States of America\\
$^{2}$Department of Physics, Southern Methodist University, 3215 Daniel Avenue, Dallas, TX 75275, USA\\
$^{3}$Department of Physics, New York University, 726 Broadway, New York, NY 10003, USA\\
$^{4}$Center for Computational Astrophysics, Flatiron Institute, 162 5th Avenue, New York, NY 10010, USA\\
$^{5}$NASA Goddard Space Flight Center, 8800 Greenbelt Road, Greenbelt, MD 20771, USA\\
$^{6}$Department of Astronomy, Cornell University, 122 Sciences Drive, Ithaca, NY 14853, USA
}

\date{Accepted XXX. Received YYY; in original form ZZZ}

\pubyear{\the\year{}}

\begin{document}
\label{firstpage}
\pagerange{\pageref{firstpage}--\pageref{lastpage}}
\maketitle

\begin{abstract}
Line intensity mapping (LIM) is a technique for producing 3D maps of the Universe by scanning the sky with a spectrometer sensitive to a range of wavelengths corresponding to the redshifted spectral lines of atoms or molecules, such as hydrogen or carbon, commonly found in galaxies and the diffuse media around them. While LIM experiments have successfully detected the 21 cm line of neutral hydrogen, other lines that reveal large-scale structure or astrophysical processes remain undetected. Many LIM experiments are in development or are underway to fill this gap, but will likely suffer from contamination from systematics, like Galactic foregrounds, or noise. Cross-correlation techniques offer the smoothest route for making detections and constraining astrophysical processes in this regime. In this work, we apply three cross-correlation techniques (stacking, the conditional voxel intensity distribution (CVID), and the cross power spectrum) to simulated LIM maps produced using [CII] luminosity models for a pathfinder LIM experiment (EXCLAIM). We find that these cross-correlation techniques allow for mean detection of the target signal line ([CII]) at redshifts 2.5-3.5 at the 4.5$\sigma$, 3.9$\sigma$, and 8.4$\sigma$ level, respectively, and offer moderate constraints on the line emission model. Under a futuristic scenario with reduced noise, the techniques improve substantially, with detections at the 44.0$\sigma$, 24.6$\sigma$, and 34.3$\sigma$ levels and percent-level constraints. Each technique offers unique information, with the strongest constraints achieved by using the three techniques in combination.
\end{abstract}




\section{Introduction}

Understanding the evolution of large-scale structure (LSS) from the time of the Big Bang to the present epoch is the primary goal of cosmologists. The observation of this evolution after the emission of the cosmic microwave background (CMB) and the formation of the first luminous objects typically depends on galaxy surveys. By their very nature, such surveys are only sensitive to galaxies bright enough to be detected individually, thereby omitting a large number of fainter galaxies that would provide useful information. Deeper surveys can also suffer from field-to-field variance due to small survey regions, as in \citet[]{popping_alma_2019}.

The formation and evolution of galaxies remain topics of particular interest in astrophysics, but are similarly hindered by the limitations of the traditional galaxy surveys. One particular aspect of this phenomenon is the star formation rate (SFR) in a galaxy \citep[]{madau_cosmic_2014}. As a galaxy begins to form, cooler gas coalesces to form stars. Older galaxies contain less of this gas and host other phenomena which suppress star formation, such as active galactic nuclei. "Cosmic noon," or the epoch of peak star formation in galaxies as an aggregate, is believed to occur around redshift $z=2-3$ \citep[]{madau_cosmic_2014}, where it is difficult for traditional galaxy surveys to accurately record the evolution for all galaxies over wide swaths of the sky.

\emph{Line intensity mapping} (LIM) is a novel technique for mapping LSS and studying galactic evolution that can overcome some of the drawbacks of traditional surveys. Instead of observing a single, small portion of the sky for an extended period of time, LIM surveys measure the surface brightness of larger sky areas relatively quickly, integrating the emissions from multiple galaxies at once, including those which would typically be too dim to appear in traditional surveys. This allows for efficient LSS mapping. Furthermore, LIM surveys target frequencies corresponding to specific atomic or molecular transitions that trace LSS or other astrophysical phenomena of interest, such as those from common galactic constituents like H, C, N, O, or CO.

Numerous experiments have begun or are in development, targeting one or more of these constituents over a variety of redshift ranges and survey areas. \citet[]{bernal_line-intensity_2022} illustrate the wide range of survey strategies in their Fig. 3. Many of these experiments are ground-based telescopes, like the CO Mapping Array Project \citep[COMAP]{cleary_comap_2022} and the Canadian Hydrogen Intensity Mapping Experiment \citep[CHIME]{chimefrb_collaboration_chime_2018}, which both operate in the radiowave portion of the electromagnetic spectrum. SPHEREx \citep[]{dore_cosmology_2015} is a recently launched satellite creating full-sky LIM maps in the near-infrared band. Still other experiments use balloons to temporarily lift telescopes above much of the atmosphere, like TIM \citep[]{vieira_terahertz_2020}, which operates in the far-infrared.

In this work, we seek to model a typical pathfinder LIM mission to demonstrate the performance of cross-correlation techniques. While our analyses can be performed on any given LIM experiment, we choose to model the EXperiment for Cryogenic Large-Aperture Intensity Mapping (EXCLAIM) as an example. The experiment's design consists of a balloon-borne telescope targeting the [CII] emission line of ionized carbon over the frequencies $420-540$ GHz, corresponding to redshifts $z=2.5-3.5$ \citep[]{switzer_experiment_2021}. This line was chosen specifically for its ability to trace the SFR at these important redshifts in galactic evolution \citep[]{de_looze_applicability_2014, herrera-camus_c_2015, pentericci_tracing_2016, aravena_alma_2016}. The EXCLAIM frequency band also coincides with a number of other emission lines coming from galaxies at different redshifts, such as the CO $J=4-7$ rotational lines, the OI and OIII lines of oxygen, and the NII lines of nitrogen \citep[]{silva_mapping_2021}. While these lines could also be used for astrophysics, [CII] is by far the brightest line and offers the best chance at an EXCLAIM detection. 

While [CII] has been tentatively detected in a cross-correlation between Planck High-Frequency Instrument (HFI) maps and BOSS quasars \citep[]{yang_evidence_2019}, doing so with EXCLAIM will require a comprehensive understanding of not just the signal, but also contaminants and instrumental noise. Contaminants include not only the line interlopers, i.e., the line emissions from CO, O$_2$, and N$_2$, but also foregrounds from the continuum emission of dust in the Milky Way or the cosmic infrared background (CIB). EXCLAIM is designed to operate close to the background photon noise limit imposed by residual atmospheric emission at ballooning altitudes through the use of a cryogenically cooled optics system containing $\mu$-spec spectrometers \citep[]{essinger-hileman_optical_2024, mirzaei_-spec_2020, volpert_developing_2022}. Nevertheless, the EXCLAIM maps will contain significant amounts of noise. The contaminants and noise therefore suggest a strong preference for cross-correlation techniques when making a detection with EXCLAIM.

In \citet[]{pullen_extragalactic_2023}, the authors forecast results of an EXCLAIM mission using a cross-power spectrum analysis and a conditional voxel intensity distribution (CVID) analysis. These techniques cross-correlate analytic models of the [CII] emission in the portion of the EXCLAIM survey area known as Stripe 82 \citep[S82]{abazajian_seventh_2009}, a region 260 deg$^2$ along the celestial equator, with the Sloan Digital Sky Survey (SDSS) Baryon Oscillation Spectroscopic Survey (BOSS) \citep[]{alam_eleventh_2015, ahumada_16th_2020}. The S82 portion of the EXCLAIM survey was specifically chosen to overlap with this region. In performing the forecast, the authors calculate analytically the contributions of the [CII] signal, the CO line interloper, instrumental noise, Galactic continuum emissions, and a finite point spread function (PSF) to mimic the instrument beam. For the cross-power spectrum, the authors achieve signal-to-noise ratios (SNRs) between $4.35-11.4$ depending on the [CII] emission model, the frequency range used, and whether the lowest $k_{\|}$ mode is removed. Similarly, for the CVID, the authors achieve SNRs of $5.8-24.5$.

In the present work, we attempt a more robust forecast of an EXCLAIM-like experiment's ability to detect [CII] emission. Using the \verb|Limlam Mocker| Python package from \citet[]{stein_georgesteinlimlam_mocker_2023} and dark matter halo catalogues simulated via the mass-Peak Patch algorithm \citep[]{stein_mass-peak_2019}, we create simulated 3D maps of the S82 EXCLAIM survey region. These maps include the [CII] signal, CO line interlopers, Galactic foregrounds, and instrumental white noise. The maps are generated using the EXCLAIM beam specifications, including a Gaussian beam in the transverse directions and frequency band-specific  noise levels.

In this work, we perform three cross correlation analyses: a stacking analysis, a CVID analysis, and a cross power spectrum analysis. Stacking (e.g., for 21 cm, \citet[]{collaboration_detection_2025} and \citet[]{chen_emission-line_2025}) and the cross power spectrum (e.g., also for 21 cm, \citet[]{anderson_low-amplitude_2018} and \citet[]{wolz_hi_2022}) have been used previously to successfully detect line emission, but the CVID \citep[]{breysse_canceling_2019} is relatively novel and has yet to be tested on LIM data. We seek to compare these techniques to one another and demonstrate their unique and combined capabilities. We perform these cross-correlations using the specifications of the quasar (QSO) survey of SDSS's BOSS and extended BOSS (eBOSS) in S82 and a simulated LIM map premised on an empirical model for [CII] emission for these galaxies contaminated by line interlopers, Galactic foregrounds, and instrumental effects/noise. We forecast signal-to-noise ratios for these techniques to demonstrate their ability to detect the underlying [CII] signal. We also estimate their separate and combined ability to constrain our fiducial [CII] emission model. Ultimately, we show that an EXCLAIM-like LIM experiment is capable of a [CII] detection and soft constraints on galactic [CII] emission. Cross-correlation is of course not limited to any one experiment, and could be applied to any of those mentioned above, or to an improved, EXCLAIM-like experiment. Under such a futuristic scenario in which noise is reduced, the ability to detect and constrain [CII] emission increases significantly.

In Sec. \ref{sec:simcomps}, we describe the simulation components and their underlying models. In Sec. \ref{sec:crosscorr}, we define the cross-correlation techniques utilized as well as the SNR and model constraint methods. In Sec. \ref{sec:results}, we present the results from these techniques as applied to the simulations. Finally, in Sec. \ref{sec:disc} we discuss the results and considerations for future analyses.

\section{Simulation Components}\label{sec:simcomps}

\begin{figure*}
    \centering
    \includegraphics[width=0.8\linewidth]{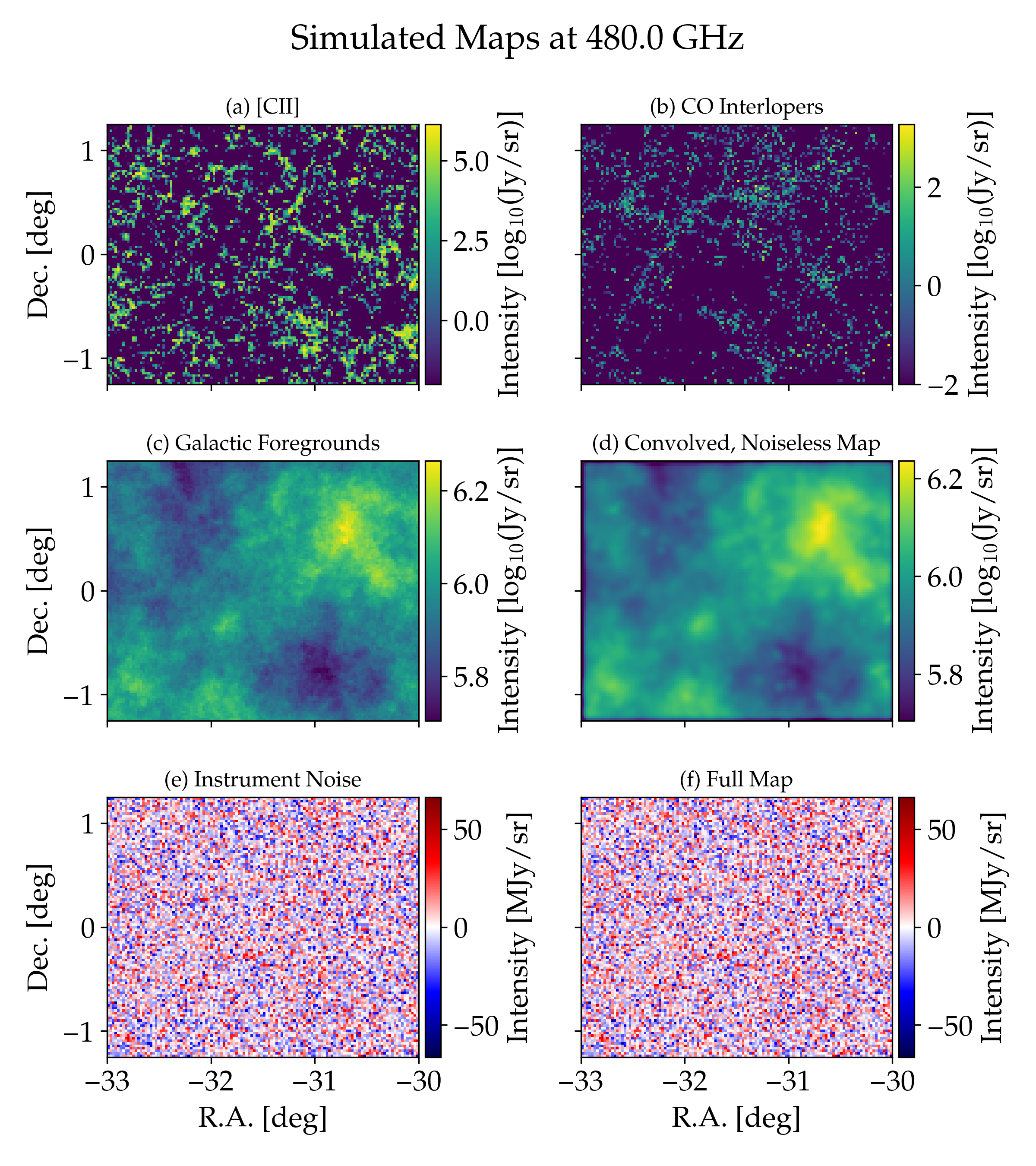}
    \caption{A patch of simulated LIM maps generated in \texttt{Limlam Mocker}, as described in Sec. \ref{sec:simcomps}. (a), (b), (c), and (e) are individual map components. (d) is the sum of (a), (b) and (c) convolved by the angular and spectral resolution of the EXCLAIM instrument. (f) is the sum of (d) and (e). Note that (a), (b), (c), and (d) are in log$_{10}$(Jy/sr) units and (e) and (f) are in MJy/sr.}
    \label{fig:patchmaps}
\end{figure*}

\subsection{[CII] Signal Emission}\label{sec:cii}

The portion of S82 planned to be surveyed by EXCLAIM is, in celestial coordinates, between $-1.25\degree<\text{Dec.}<1.25\degree$ and $-45\degree<\text{R.A.}<60\degree$, for a total sky coverage of 262.5 deg$^2$. Our [CII] signal, at 420-540 GHz, comes from redshifts 2.5-3.5. At these redshifts, we choose to model the [CII] emission from a given dark matter halo according to the \citet[]{padmanabhan_constraining_2019} model, given by:
\begin{equation}
    L_{\text{[CII]}}(M,z)=\left(\frac{M}{M_1}\right)^{\beta}\exp(-N_1/M)\left(\frac{(1+z)^{2.7}}{1+\left(\frac{1+z}{2.9}\right)^{5.6}}\right)^{\alpha},
\end{equation}
where $L$ is the luminosity of [CII] in solar luminosities, $M$ is the mass of the dark matter halo in solar masses, $z$ is the redshift of the halo, and $\{\alpha, \beta, M_1, N_1\}=\{1.79, 0.49, 2.39\times 10^{-5} \text{ M}_{\odot}, 4.19\times 10^{11}\text{ M}_{\odot}\}$ are the model parameters. The given values for these parameters are the best fit values found in \citet[]{padmanabhan_constraining_2019}.

We choose this model to match the analysis in \citet[]{pullen_extragalactic_2023}. As those authors note, there are many models for [CII] emission and they vary by orders of magnitude in their predictions. The diversity of models likely comes from the unique ways in which they are developed. Some are based on attempts to measure [CII] emission from individual galaxies directly, while others are based on detailed simulations of galactic processes. The \citet[]{padmanabhan_constraining_2019} model is one of the more optimistic models, but it is calibrated to a more recent [CII] measurement \citep[]{yang_evidence_2019} than other models.

To generate [CII] intensity maps from this model, we use mass-Peak Patch dark matter halo simulations. In the mass-Peak Patch method, an initial, 3D overdensity field is generated by convolving a white noise field with the linear matter power spectrum for a given set of cosmological parameters. Dark matter halo candidates are found by smoothing the density field at varying scales and identifying locations where there is an overdensity above a specified threshold. The mass of the halo is assigned by determining whether the overdense region will collapse at the given redshift. Final velocities and locations are assigned according to the displacement fields and growth factors. For a detailed description on how these simulations work, see \citet[]{stein_mass-peak_2019}.

Ultimately, our simulations produced maps of dark matter halos for redshifts 2.5-3.5 with a minimum halo mass of $\sim 3\times 10^{10}\,M_\odot$. Each map is broken into 3 deg$\times$ 3 deg (R.A. $\times$ Dec.) patches, which we trim to 3 deg.$\times$ 2.5 deg. to match the dimensions of the survey region. 35 of these patches arranged along the R.A. direction are therefore sufficient to cover the survey region. Note that, due to simulation limitations, the entire S82 survey region cannot be simulated at once as a single strip, so "adjacent" patches might be from different simulations and are not necessarily correlated with one another with respect to large scale structure. We take care in our cross-correlations to avoid using the edges of the patches and to refrain from drawing conclusions about scales larger than those found within a single patch (see Sec. \ref{sec:crosspower}).

From a patch, a catalogue is made of the mass and location of each halo in comoving coordinates. To convert this into a simulated EXCLAIM map, we use \verb|Limlam Mocker|\footnote{\url{https://github.com/georgestein/limlam_mocker}}. This package reads in the halo catalogue and produces a mock LIM map according to the provided luminosity model.

Halo luminosities are converted into intensities in units of Jy/sr, with each voxel containing the sum intensity of all dark matter halos contained within its bounds. Here, we use \textit{voxel} to mean a 3D bin of the map, i.e., a pixel with a depth component. \verb|Limlam Mocker| allows for customizing the voxelization of the LIM map. The size of our voxels, determined by the resolution of the EXCLAIM instrument (see Sec. \ref{sec:resolution}), are $0.024\degree\times 0.024\degree\times0.36$ GHz, meaning a single patch is $125\times104\times333$ voxels large. An example of a simulated patch of [CII] is shown in Fig. \ref{fig:patchmaps}(a).

\subsection{Line Interlopers}

A \textit{line interloper} is an emission line other than [CII] of some atom or molecule that exists at a different comoving distance but happens to be redshifted to the same frequencies at which we observe [CII], i.e., it is a contaminant of the signal. At 420-540 GHz, [CII] is expected to be the brightest emission line by orders of magnitude \citep[]{silva_mapping_2021}, but line interlopers include rotational lines from CO and other transition lines from N and O. In their analysis, \citet[]{pullen_extragalactic_2023} consider the CO rotational lines, emitted at redshifts of $<1$ as another target signal, but here we treat CO solely as a contaminant.

For this work, we simulate only the CO emission in our maps. Given that nearly every line interloper is of equal brightness and how bright [CII] is relative to them \citep{silva_mapping_2021}, this choice preserves the presence of interloping structure while reducing the need for halo catalogues at wider redshift ranges without dimming the contamination to an unreasonable extent.

To simulate the CO emission, we use a similar process to that of [CII]. Peak Patch catalogues for redshifts 0.1-1 are loaded into \verb|Limlam Mocker| and intensity maps are produced using the \citet[]{li_connecting_2016} and \citet[]{keating_intensity_2020} CO model, following the choice made by \citet[]{pullen_extragalactic_2023}. Specifically, we simulate the CO rotational lines for $J=7-6,\:6-5,\:5-4,\:4-3$ and the Li-Keating model can be written as:
\begin{equation}
    L'_{\text{CO}}(J)=\frac{\log(L_{\text{IR}})-\beta_{\text{CO}}(J)}{\alpha_{\text{CO}}(J)},
\end{equation}
where $L'_{\text{CO}}$ is the CO luminosity in K km s$^{-1}$ pc$^2$, $L_{\text{IR}}$ is the infrared luminosity associated with the halo and determined by the luminosity-star formation rate relationship described by \citet[]{robert_c_kennicutt_global_1998} and \citet[]{carilli_cool_2013}, and $\alpha_{\text{CO}}$ and $\beta_{\text{CO}}$ are parameters given in \citet[]{kamenetzky_lcolfir_2016}. The star formation rate can be calculated from the dark matter halo mass using the relations found in \citet[]{behroozi_average_2013}. An example of a simulated patch of the CO interlopers is shown in Fig. \ref{fig:patchmaps}(b). As expected, the interlopers are largely dimmer than the [CII] signal.

\subsection{Galactic Foregrounds}\label{sec:foregrounds}

While S82 is off the Galactic plane, thermal dust emission from the Milky Way remains a strong contaminant at our frequencies. Our simulations include foregrounds from the Milky Way simulated using the \verb|PySM| Python package \citep[]{Panexp_2025, Zonca_2021, Thorne_2017}. We use the \verb|d1| thermal dust model for the S82 sky region with an $N_{\text{sides}}=512$ in each frequency/redshift bin of the simulated map. Other components of the Galactic foregrounds, like free-free emission, are omitted since they are much dimmer at these frequencies and in these sky regions. An example of a patch of the foregrounds is shown in Fig. \ref{fig:patchmaps}(c). Despite choosing a sky region with relatively dim foregrounds, the foregrounds are considerably brighter than the underlying [CII] signal, often by an order of magnitude or more. The Galactic foregrounds are therefore the largest, non-instrumental signal contaminant. For a "perfect" instrument, our LIM maps would consist of the [CII], CO, and Galactic foregrounds maps added together.

\subsection{Instrument Resolution}\label{sec:resolution}

So far, the simulation components have been entirely independent of the observing instrument. For our analysis, we are simulating LIM maps as observed by the EXCLAIM instrument, or one similar to it. EXCLAIM seeks to reduce thermal noise and atmospheric background contamination by submerging the telescope in a liquid helium-filled dewar and elevating it using a balloon to 36 km above the Earth's surface. The telescope itself is a 75 cm aperture two-mirror Gregorian telescope that will ultimately focus light on six $\mu$-Spec integrated spectrometers \citep[]{volpert_developing_2022}. On each spectrometer, there are 355 microwave kinetic inductance detectors (MKIDS), each corresponding to a different frequency band. Of these, 333 will be used across the 420-540 GHz band for producing LIM maps (the remainder are left for calibration). For a more detailed summary of EXCLAIM, see \citet[]{pullen_extragalactic_2023}. For the full technical description, see \citet[]{switzer_experiment_2021}.

The EXCLAIM instrument has both an angular and spectral resolution by which we convolve our simulated maps \citep[]{switzer_experiment_2021}. Like \citet[]{pullen_extragalactic_2023}, we use a 4.33 arcminute full width half maximum for a Gaussian point spread function in the angular plane. While the EXCLAIM resolution varies over the frequency range (between 4.86' and 3.78'), we choose to maintain a constant resolution at the middle of this range for simplicity. With the voxelization described below, omitting the variation in the resolution would likely not produce any notable effects as the bounds of the EXCLAIM resolution are less than one voxel apart. A constant resolution in the middle of the range will therefore capture both the moderately worse SNR at lower resolution and the better SNR at higher resolution. In the frequency direction, we use a sinc$^2$ spectral resolution function with first order nulls separated by $\delta\nu=\nu/R$, where $\nu$ is the frequency and $R$ is the resolution. Again, while the resolution varies over the frequency range (between 535 and 505), we choose a constant $\delta\nu=472\text{ GHz} / 512$ for simplicity.

The voxelization of the simulated maps is determined by the resolution. We choose to oversample the beam with three voxels per FWHM in the angular plane. Similarly, there are three voxels per $\delta\nu$ in the frequency direction, which matches the number of detectors usable for data collection over the full frequency range. A convolved map is shown in Fig. \ref{fig:patchmaps}(d). Note that the convolution is performed on each patch separately, leading to a border of voxels at the edge of the patch maps where the intensity is notably dimmer.

\subsection{Instrument Noise}

The final component added to the simulated EXCLAIM maps is instrumental thermal noise. EXCLAIM is cryogenic and balloon-borne, so thermal noise and noise from the atmosphere are suppressed, but still pose the largest threat to detectability. Additionally, artifacts of the scan strategy and $1/f$ noise can produce correlations across the map unattributable to the [CII] signal. The impact of noise is highlighted by \citet[]{pullen_extragalactic_2023} as one of the primary reasons for using cross-correlation techniques for EXCLAIM.

For this work, we do not include $1/f$ noise because the stacking and CVID techniques described below retrieve small-scale information and are thought to be robust against such large-scale effects. We leave consideration of more complicated noise models and other instrumental effects to future work. This leaves a Gaussian noise component with a unique noise-equivalent intensity (NEI) in each frequency bin \citep[]{switzer_experiment_2021}. We assume that S82 is observed for 10.5 hours divided equally across every pixel. The NEI, scaled by the observing time to intensity in units of Jy/sr, serves as the standard deviation of the Gaussian, from which a noise intensity is drawn for each voxel in a given frequency bin. Atmospheric lines cause the NEI to vary by up to an order of magnitude across the frequency range. An example of a noise map patch is shown in Fig. \ref{fig:patchmaps}(e). The noise level reaches up to $\sim$50 MJy/sr at this frequency, making it brighter than the [CII] signal by two orders of magnitude. In Fig. \ref{fig:patchmaps}(f), we see a full simulated map patch, consisting of the convolved [CII], CO, and foregrounds added to the instrument noise. The [CII] is indistinguishable by eye.

\subsection{Quasars}\label{sec:quasars}
To perform the cross-correlations detailed below, we generate mock catalogs of QSOs similar to those produced by (e)BOSS. To generate the QSO catalogue, halos within the simulation with a mass exceeding $4\times10^{12}\;M_{\odot}$ are flagged as being a potential QSO host. This threshold was chosen based on \citet[]{rodriguez-torres_clustering_2017}, which suggests QSOs identified by BOSS at high redshifts inhabit halos with a mass greater than this level. The simulated EXCLAIM survey region contains $\sim$120,000 halos that meet this criterion.

A subset of the flagged halos are chosen at random with uniform probability to host QSOs. For the one-point statistics (stacking and the CVID), 4,464 halos are chosen to match the number of QSOs found in the eBOSS catalog in the EXCLAIM survey region.\footnote{The halo numbers used here were determined using the searchable SQL DR17 (e)BOSS database found at \url{https://skyserver.sdss.org/dr17/SearchTools/sql\#}.} From these, 3,328 halos are chosen for the two-point statistic (cross power spectrum) to match the number of QSOs in the BOSS catalog in the same region. The inhomogeneous sampling of eBOSS compared to the BOSS subset restricts its use to just the 1pt statistics, but we do not attempt to model the exact angular or redshift distribution of the (e)BOSS QSOs. We assume that halos can contain only one QSO each.

\section{Cross-Correlation Methods}
\label{sec:crosscorr}

In this work, we perform three different cross-correlation analyses. There are two one-point statistics, the more traditional stacking analysis and the more novel conditional voxel intensity distribution, and one two-point statistic, the cross power spectrum. In the following subsections, we detail the method behind each technique, as well as how they can be used to constrain the [CII] intensity model. All of the cross-correlations are performed between the simulated, EXCLAIM-like intensity maps and the mock, (e)BOSS-like QSO catalogue maps. 

\subsection{Stacking}
\label{sec:stacking}

Our stacking analysis is inspired by that found in \citet[]{DunneEtAl2023}, with some modifications adopted for the specifics of an EXCLAIM-like experiment.

To create a stack, the mock QSO catalog map is used to identify voxels in the simulated intensity map which would host a QSO. For each of these voxels, a subsection of the map centered on the QSO-containing voxel, called a \textit{cubelet}, is isolated. This cubelet is a 3-dimensional object, consisting of adjacent voxels in both the angular directions, as well as the frequency dimension. Each voxel in the map is also assigned an error value based on its frequency slice's observation time-scaled NEI value. Therefore, for each cubelet created, a corresponding error cubelet is also created.

Once all cubelets are created, they are each rotated 0$\degree$, 90$\degree$, 180$\degree$, or 270$\degree$ in the R.A.$\times$Dec. plane to reduce potential bias or asymmetry, such as from the scan strategy, that might arise in the LIM observations. For each cubelet, we also subtract the mean intensity value per frequency slice in order to reduce the impact of the galactic foregrounds on the stacking measurement. Such foregrounds are slowly varying over the scale of a cubelet in the angular dimensions ($\sim$3 voxels, or $\sim0.072\degree$), so mean subtraction acts as a rudimentary foreground removal process.

After this pre-processing of the cubelets, they are co-added voxel-wise, or \textit{stacked}, using inverse variance weighting (IVW). If a given cubelet $i$ has a voxel intensity $I_i$ with error $\sigma_i$, the corresponding weighted mean intensity $\hat{I}$ of all cubelets' voxels would be given by
\begin{equation}
    \hat{I}=\frac{\sum_i I_i / \sigma_i^2}{\sum_i 1 / \sigma_i^2}
\end{equation}
and has error
\begin{equation}
    \sigma_I^2 = \frac{1}{\sum_i 1 / \sigma_i^2}.
\end{equation}

The stack is a 3D object that approximately maps the mean [CII] emission around a mock QSO. For further analysis, we condense this information into a single number by adding the intensities of the central voxel and the voxels immediately adjacent together. These voxels are chosen as they represent the majority of the signal given the beam's FWHM ($\sim$3 voxels) and spectral resolution ($\sim$ 3 voxels). Adjacent voxels can also contain some [CII] signal from non-QSO host halos within a local cluster (see, e.g., \citet[]{dunne_three-dimensional_2025}). In summing these central twenty-seven voxels, we weight them according to the beam's angular resolution and spectral response. The final sum is an approximate measurement of the total mean [CII] emission from the vicinity of a mock QSO.

To calculate a signal-to-noise ratio, we compare the QSO stack to stacks performed on random locations on the intensity map. We generate 1,000 instances of the instrument noise to overlay on the intensity map, and for each instance, we select a new, random set of QSOs on which to stack.\footnote{In each instance, the same set of halo catalogues are used. While it would be possible to also vary the set of catalogues, it is more computationally expensive and we notice no significant changes in the results when doing so.} For the same instance, we select random locations on the intensity map (equal in number to the number of QSOs) and perform the same stacking process on them. This is done 250 times for each instance, creating a distribution of randomly-centered stacks. After summing the central 27 voxels in the QSO stack and random stacks in the same manner described above, we calculate the SNR as
\begin{equation}\label{eq:stacking_snr}
    \text{SNR} = \frac{I_{\text{QSO}}-\bar{I}_{\text{random}}}{\sigma_{I_{\text{random}}}},
\end{equation}
where $I_{\text{QSO}}$ is the intensity of the QSO stack, $\bar{I}_{\text{random}}$ is the mean intensity of the random stacks, and $\sigma_{I_{\text{random}}}$ is the standard deviation of the random stack intensities. An SNR is calculated the same way for each noise instance, creating a distribution of SNRs from which we report our results.

\subsection{Conditional Voxel Intensity Distribution}
\label{sec:cvid}

The conditional voxel intensity distribution (CVID) is a one-point statistic developed in \citet[BREYSSE]{breysse_canceling_2019} that will be adapted to this analysis. It was applied in a general, semi-analytic manner to EXCLAIM in \citet[]{pullen_extragalactic_2023}, but here we apply it with the full simulations.

If one considers a single voxel in the intensity map, its intensity value will be dependent on $N$, the number of [CII]-emitting halos within its volume, and the intensity $I$ that $N$ halos collectively emit. Due to the variability in the underlying halo masses, we can write an expression for the probability a voxel will have a given intensity, i.e., the \textit{voxel intensity distribution} (VID):
\begin{equation}
    \mathcal{P}_{\text{[CII]}}(I)=\sum_{N=0}^{\infty} \mathcal{P}_N(I)\mathcal{P}(N),
\end{equation}
where $\mathcal{P}_N(I)$ is the probability a voxel containing $N$ [CII]-emitting halos has intensity $I$, and $\mathcal{P}(N)$ is the probability a voxel contains $N$ [CII]-emitting halos. The intensity maps contain not only [CII], but also interlopers, foregrounds, and instrument noise, so one can think of the [CII] VID as being convolved with VIDs from the contaminants:
\begin{equation}\label{eq:vid}
    \mathcal{P}(I)=\mathcal{P}_{\text{[CII]}}\circ\mathcal{P}_{\text{cont}}(I).
\end{equation}
Here, we assume that these contaminants have no correlation with the [CII] signal.

To remove the contaminants and probe the signal more directly, we utilize the \textit{conditional voxel intensity distribution} (CVID). Rather than creating our probability distributions using all voxels in the map, we divide them into two sets: voxels containing at least one QSO ($N_{\text{det}}\geq1$), and the rest ($N_{\text{det}}=0$). Eq. \ref{eq:vid} can be rewritten for the CVID as
\begin{equation}\label{eq:cvid}
    \prob{}{}{I}{} = \prob{}{[CII]}{I}{} \circ \mathcal{P}_{\text{cont}}(I).
\end{equation}
Because the QSOs are expected to be correlated with the [CII] emission, but not the contaminants, the contaminant VID is identical between the two sets of voxels. We can exploit this fact by taking the Fourier transform of Eq. \ref{eq:cvid}:
\begin{equation}\label{eq:ft_cvid}
    \ftprob{}{}{\tilde{I}}{}=\ftprob{}{[CII]}{\tilde{I}}{}\cdot\tilde{\mathcal{P}}_{\text{cont}}(\tilde{I}),
\end{equation}
where $\tilde{I}$ is the Fourier conjugate of $I$ and
\begin{equation}
    \tilde{\mathcal{P}}(\tilde{I})=\int_{-\infty}^{\infty}\mathcal{P}(I)e^{iI\tilde{I}}dI.
\end{equation}

The Fourier transform changes the convolutions into multiplications, allowing for the uncorrelated contaminant fields to be divided out like:
\begin{align}\label{eq:ratio}
    \mathcal{R} \equiv \frac{\ftprob{}{}{\tilde{I}}{=0}}{\ftprob{}{}{\tilde{I}}{=1}} &= \frac{\ftprob{}{[CII]}{\tilde{I}}{=0}\cdot\tilde{\mathcal{P}}_{\text{cont}}(\tilde{I})}{\ftprob{}{[CII]}{\tilde{I}}{=1}\cdot\tilde{\mathcal{P}}_{\text{cont}}(\tilde{I})} \\
    &=\frac{\ftprob{}{[CII]}{\tilde{I}}{=0}}{\ftprob{}{[CII]}{\tilde{I}}{=1}}.
\end{align}
We refer to $\mathcal{R}$ as the (Fourier-transformed) CVID ratio.

The probability distributions in the VID and CVID are, in theory, continuous, but must be approximated by constructing histograms of the voxels from the intensity map. For an intensity bin $i$ centered at $I_i$ with sufficiently narrow width $\Delta I$, the number of voxels in said bin $B_i$ is
\begin{equation}\label{eq:Bi}
    B_i = N_{\text{vox}} \int_{I_i - \Delta I/2}^{I_i + \Delta I/2} \mathcal{P}(I) dI \approx \mathcal{P}(I_i)N_{\text{vox}}\Delta I,
\end{equation}
where $N_{\text{vox}}$ is the total number of voxels contributing to the VID/CVID estimation. Thus, one can estimate $\mathcal{R}$ using this linear relationship.

While the noise within a given frequency slice is Gaussian, the intensity distribution of voxels across all frequency slices is non-Gaussian given the varying NEI across frequency. Therefore, to perform the calculation of $\mathcal{R}$ using the simulated intensity map, we first inverse weight all voxel intensities by their associated error\footnote{One alternative method would be to perform the CVID ratio calculation within each frequency bin separately and combine the results at the end. Exploring the nuances of these choices is left to future work.} to produce the dimensionless intensity $\hat{I}$:
\begin{equation}\label{eq:noiseweight}
    \hat{I}_i = \frac{I_i}{\sigma_i}.
\end{equation}
Voxels are then separately histogrammed according to whether they contain at least one QSO according to the mock catalogue. We include voxels directly adjacent to those containing QSOs in the $N_{\text{det}}\geq1$ set of histogrammed voxels. This mimics the stacking process, in which adjacent voxels are included in the summed intensity. All other voxels are included in the $N_{\text{det}}=0$ histogram. The binning for the histograms is set such that there are $\sim$40,000 bins between -5.69 and 5.69 (weighted intensity). The number of bins was chosen to ensure they are sufficiently narrow to (a) meet the assumption of Eq. \ref{eq:Bi}, and (b) allow the signal, which is about 2-3 orders of magnitude dimmer than the noise, to still be distinguishable across many bins in the absence of noise. The range of bins was chosen to permit adequate mode sampling of the histograms. 

The histograms are Fourier transformed and the ratio $\mathcal{R}$ is calculated. Note that the Fourier transformed CVIDs and their ratio have both real and imaginary components. We use only the results for the $\tilde{I}>0$ bins, as the Fourier transform of the CVID produces an even (real) or odd (imaginary) function; the $\tilde{I}<0$ bins contain no additional, useful information.

The SNR of the CVID ratio can be calculated using
\begin{equation}\label{eq:cvidsnr}
    \text{SNR}^2 = [\vec{\mathcal{R}}-(1|0i)]^T \textbf{C}^{-1} [\vec{\mathcal{R}}-(1|0i)].
\end{equation}
Here, we construct $\vec{\mathcal{R}}$ from $\mathcal{R}$ by separating the real and imaginary parts in each $\tilde{I}$ bin and concatenating them as the vector components (e.g., if $\mathcal{R}$ has three bins, $\vec{\mathcal{R}}$ has six components: the three real parts concatenated with the three imaginary parts). From each component, we subtract 1 (real) or 0$i$ (imaginary) because the ratio $\mathcal{R}$ would equal $1+0i$ if it were purely noise. $\textbf{C}^{-1}$ is the covariance matrix of $\vec{\mathcal{R}}$. 

We estimate the mean CVID SNR by creating 1,000 instances of $\vec{\mathcal{R}}$, each calculated from an intensity map with a newly generated instrument noise instance and a new, randomly-selected set of QSOs. The covariance of the $\vec{\mathcal{R}}$ instances is used as $\textbf{C}^{-1}$ in Eq. \ref{eq:cvidsnr}, with each instance of $\vec{\mathcal{R}}$ producing an SNR value. The result over all instances is a distribution of SNR values.

\subsection{Cross Power Spectrum}
\label{sec:crosspower}

We calculate the cross power spectrum using a generalization of \textsc{Limlam-Mocker}'s auto power spectrum calculator, which computes the 3D power spectrum on a grid before averaging over the number of modes to create the 1D $P(k)$, where the $k$ binning is limited by the size of the map and the size of an individual voxel. The input fields are
\begin{equation}
    \delta_I(\textbf{r})=\hat{I}(\textbf{r})-\bar{I},
\end{equation}
where $\hat{I}(\textbf{r})$ is the (noise-weighted according to Eq. \ref{eq:noiseweight}) line intensity map and $\bar{I}$ is the mean voxel intensity, and
\begin{equation}
    \delta_{\text{QSO}}(\textbf{r})=\frac{N_{\text{QSO}}(\textbf{r}) - \bar{N}_{\text{QSO}}}{\bar{N}_{\text{QSO}}},
\end{equation}
where $N_{\text{QSO}}(\textbf{r})$ is a map of the number of QSOs contained within each voxel and $\bar{N}_{\text{QSO}}$ is the mean number of QSOs within a voxel. This latter field is created using the mock QSO catalogue, but with the more limited number of QSOs to reflect BOSS's smaller catalogue (see Sec. \ref{sec:quasars}). The QSOs are histogrammed according to the same voxel binning as the line intensity map. We use the noise-weighted intensity map both to remain consistent with the also noise-weighted stacking and CVID analyses and to reduce the overall noise on the cross power spectrum.


Instrument calibration drifts and $1/f$ noise can result in spurious correlations at large scales. To mimic avoiding such correlations, which do not exist in our simulations, and for computational ease, we do not compute the cross power spectrum across the entire survey region. Instead, the cross power is computed in $3\degree\times2.5\degree$ (R.A.$\times$Dec.) patches, matching the simulation patches detailed in Sec. \ref{sec:cii}. The cross power spectra of the patches are then averaged together to create one, full-map instance of the cross power.


To calculate the SNR of the cross power spectrum, we use the expression
\begin{equation}\label{eq:cross_snr}
    \text{SNR}^2=\vec{P}^T\textbf{C}^{-1}\vec{P}
\end{equation}
where $\vec{P}$ is the vectorized cross power spectrum such that each component is the value of the cross power spectrum in each $k$ bin and $\textbf{C}$ is the covariance of the cross power spectra. $\textbf{C}$ is calculated from 1,000 instances of the line intensity and QSO maps, with a new noise instance and mock catalogue, respectively. The SNR is calculated for each cross power instance, creating a distribution of SNRs.

\subsection{Constraining Intensity Model}
\label{sec:constrainA}

To demonstrate the capabilities of an EXCLAIM-like instrument in constraining [CII] emission, we consider a simple variant of the \citet[]{padmanabhan_constraining_2019} model $L_{\text{[CII]},0}(M,z)$ scaled by an amplitude factor $A$:
\begin{equation}
    L_{\text{[CII]}}(M,z)=A\:L_{\text{[CII]},0}(M,z).
\end{equation}
We use the cross-correlation techniques outlined above, both individually and in combination, to constrain this parameter by computing the probability a given value of $A$ could produce the intensity map simulated with a fiducial value of $A=1$. We use Bayesian statistics and assume the results are normally distributed.

The (natural) log probability $\mathcal{P}'(A|\textbf{x})$ can be written:
\begin{equation}
    \mathcal{P}'(A|\textbf{x})\propto \mathcal{P}'(A) + \mathcal{P}'(\textbf{x}|A),
\end{equation}
where $\mathcal{P}'(A)$ is the log prior, which we define to be a uniform distribution for $0\lessapprox A\leq 10$ and zero elsewhere. $\mathcal{P}'(\textbf{x}|A)$ is the log likelihood, for which we use:
\begin{equation}
    \mathcal{P}'(\textbf{x}|A) = -\frac{1}{2}(\textbf{x}_{\text{measured}}-\textbf{x}_{\text{model}})\textbf{C}^{-1}(\textbf{x}_{\text{measured}}-\textbf{x}_{\text{model}}).
\end{equation}
In these expressions, $\textbf{x}$ is the cross-correlation observable(s), and $\textbf{C}$ is its attendant covariance. $\textbf{x}_{\text{measured}}$ is calculated using the line intensity simulations where $A=1$, and $\textbf{x}_{\text{model}}$ is the expectation value for $\textbf{x}_{\text{measured}}$ for a given $A$ value. These expectation values are calculated using simulations without instrument noise.

For stacking, $\textbf{x}$ is the weighted sum of a stack's central 27 voxels described in Sec. \ref{sec:stacking}. For the CVID, $\textbf{x}$ is the CVID ratio described in Sec. \ref{sec:cvid}. For the cross power spectrum, $\textbf{x}$ is simply the cross power spectrum itself, as described in Sec. \ref{sec:crosspower}. The covariance $\textbf{C}$ for each is the covariance across simulation instances.

Using these statistics, we compute the log probability for a range of $A$ values within the prior's non-zero bounds. After normalizing this distribution, we determine the maximum probability value for $A$ and a 95\% confidence interval. The confidence interval is calculated by choosing the highest probability for which all values of $A$ with a greater probability have a summed probability of 95\%. The values of $A$ at the ends of this range are the lower and upper bounds.\footnote{We define the confidence interval this way so as to be agnostic with regards to the underlying probability distribution. In our simulations, it is largely Gaussian, but asymmetric distributions could also be possible. Multi-modal distributions would require more consideration.}

When constraining $A$ with multiple techniques simultaneously, we compute the three observables from the same noise instances and QSO catalogues. The observables from each instance are concatenated into one $\textbf{x}$, and we compute the covariance from the set of all such $\textbf{x}$. For simplicity, we perform the constraint calculations once with the mean $\textbf{x}$ observable as $\textbf{x}_{\text{measured}}$. We compute the maximum probability value for $A$ and the confidence interval in the same way as above.

\section{Results}\label{sec:results}

Our results are summarized in Tables \ref{tab:SNRs} and \ref{tab:A constraints}. We find that the three cross-correlation techniques would be able to detect [CII], with the cross power spectrum offering the strongest detection. Using the three techniques together provides these strongest constraints on the underlying emission model. Below, we provide more detail and commentary on these results.

\begin{table*}
    \centering
    \begin{tabular}{cccc}\hline
        \multicolumn{4}{c}{Signal-to-Noise Ratios}\\\hline\hline
        \multicolumn{4}{c}{EXCLAIM-like Noise}\\
         &    2.5th Percentile&Mean&97.5th Percentile\\\hline
         Stacking&2.51&4.47&6.63\\
        CVID&2.32&3.92&5.69\\
        Cross Power Spectrum&6.89&8.42&9.99\\\hline
        \multicolumn{4}{c}{Reduced EXCLAIM-like Noise}\\
         &2.5th Percentile&Mean&97.5th Percentile\\\hline
         Stacking&39.77&44.01&49.12\\
        CVID&22.54&24.62&26.82\\
        Cross Power Spectrum&32.30&34.26&36.18\\\hline
    \end{tabular}
    \caption{The SNRs computed as described in Secs. \ref{sec:crosscorr} and \ref{sec:results}.}
    \label{tab:SNRs}
\end{table*}

\begin{table*}
    \centering
    \begin{tabular}{ccccc}\hline
        \multicolumn{5}{c}{Constraints on $A$}\\\hline\hline
        \multicolumn{5}{c}{EXCLAIM-like Noise}\\
         &Lower Bound&Most Probable&Upper Bound&Significance ($\sigma$)\\\hline
        Stacking&0.49&0.99&1.50&3.88\\
        CVID&0.46&0.99&1.52&3.66\\
        Cross Power Spectrum&0.53&1.00&1.47&4.17\\
        Stacking + CVID&0.65&0.99&1.33&5.70\\
        Stacking + Cross Power Spectrum&0.67&1.00&1.32&6.00\\
        CVID + Cross Power Spectrum&0.66&0.99&1.33&5.90\\
        All&0.74&1.00&1.27&7.27\\\hline
        \multicolumn{5}{c}{Reduced EXCLAIM-like Noise}\\
         &Lower Bound&Most Probable&Upper Bound&Significance ($\sigma$)\\\hline
        Stacking&0.95&1.00&1.06&35.38\\
        CVID&0.97&0.99&1.02&76.64\\
        Cross Power Spectrum&0.95&1.00&1.05&38.82\\
        Stacking + CVID&0.97&0.99&1.02&88.28\\
        Stacking + Cross Power Spectrum&0.96&1.00&1.04&52.57\\
        CVID + Cross Power Spectrum&0.97&0.99&1.02&95.55\\
        All&0.97&1.00&1.01&95.55\\\hline
    \end{tabular}
    \caption{The constraining power computed as described in Secs. \ref{sec:crosscorr} and \ref{sec:results}. The probability of $A$ being between the lower and upper bounds is 95\%. The significance is the number of standard deviations between the most probable value of $A$ and 0.}
    \label{tab:A constraints}
\end{table*}

\subsection{Cross Power Spectrum Results}\label{sec:crosspowerresults}

\begin{figure*}
    \centering
    \begin{subfigure}[t]{0.49\textwidth}
        \centering
        \includegraphics[width=\textwidth]{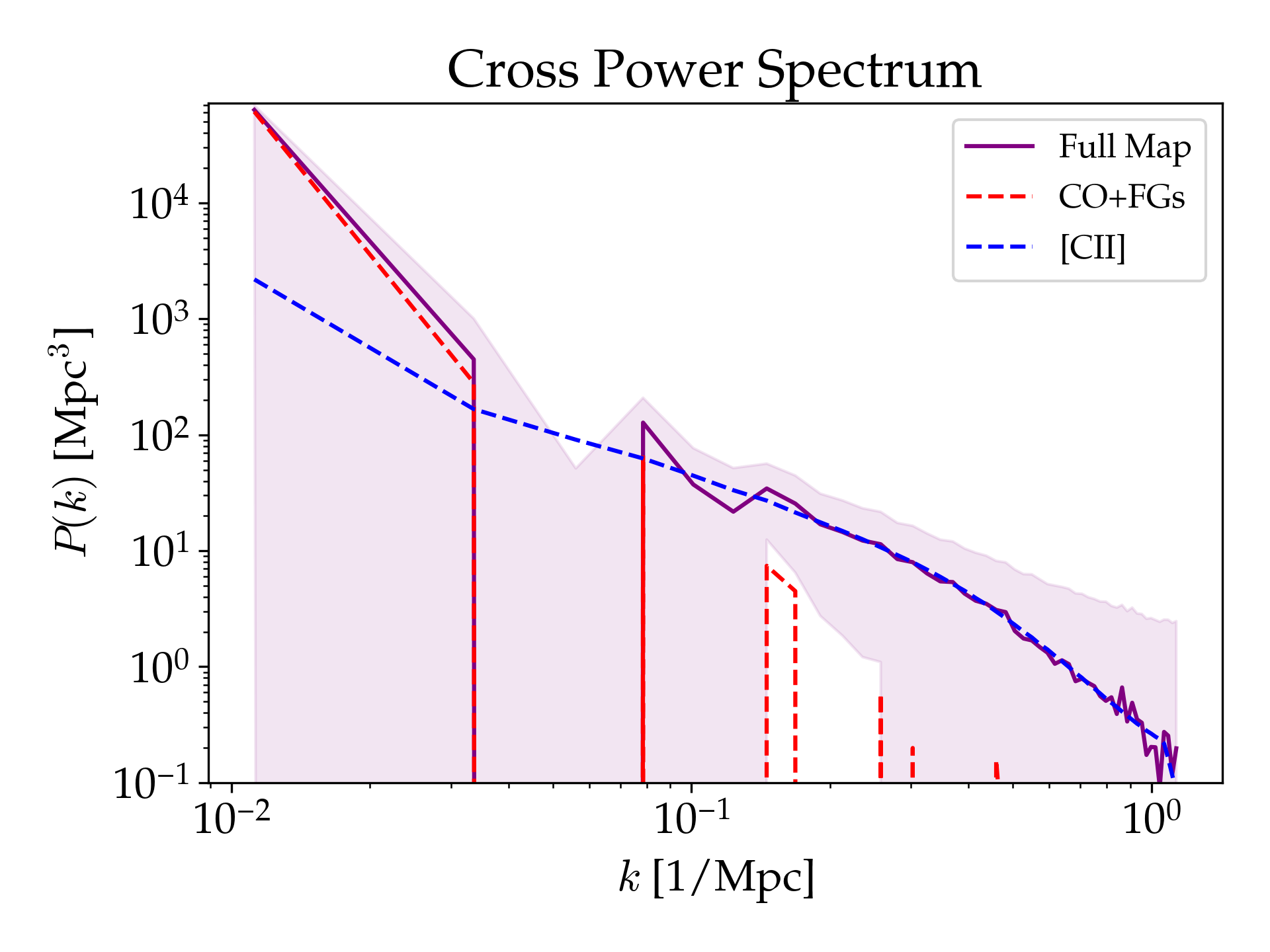}
        \caption{Unfiltered}
    \end{subfigure}%
    ~ 
    \begin{subfigure}[t]{0.49\textwidth}
        \centering
        \includegraphics[width=\textwidth]{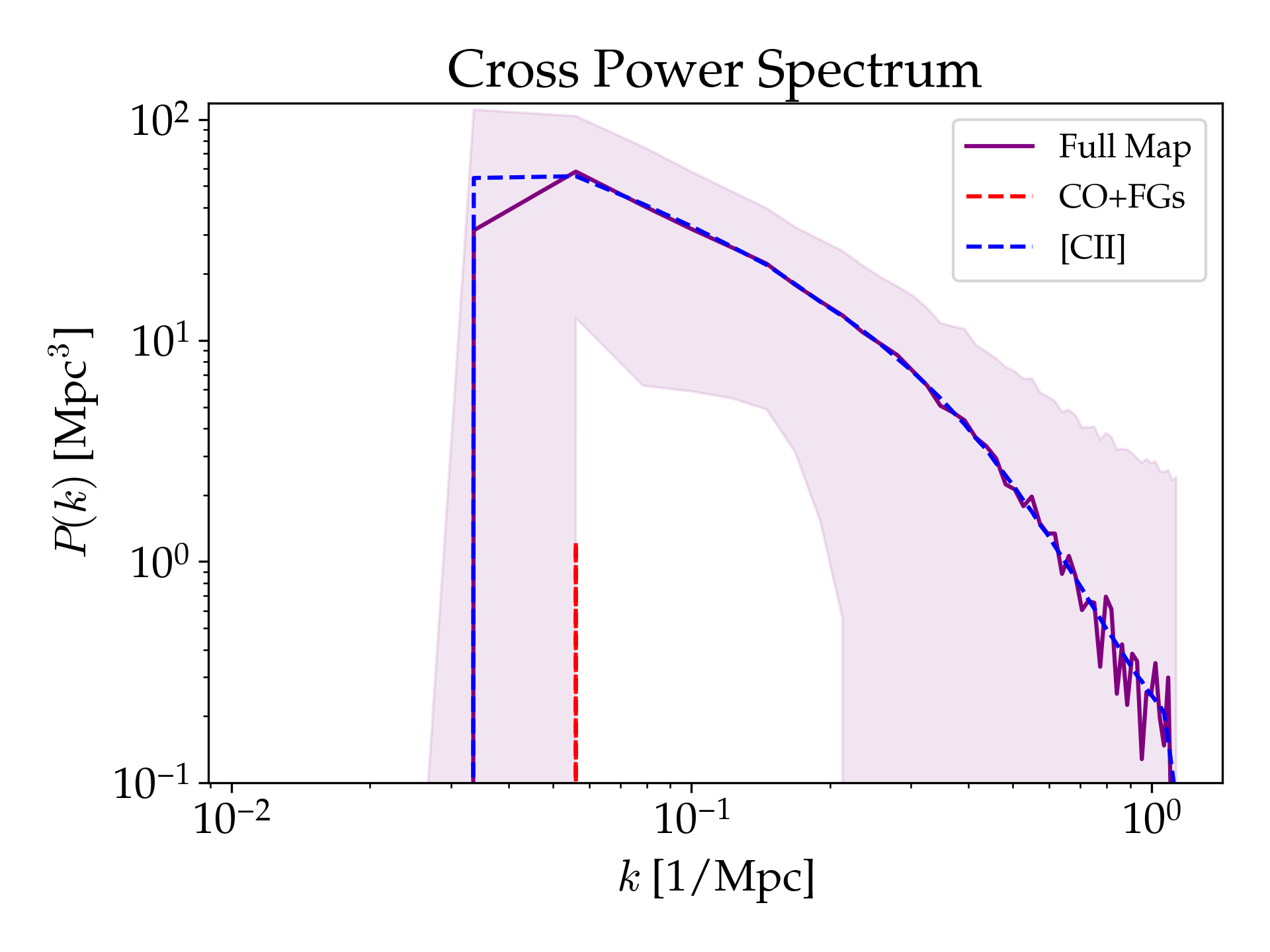}
        \caption{Filtered}
    \end{subfigure}
    \caption{The cross power spectrum between the mock QSO catalogs and simulated LIM maps. The solid purple curve is the cross power for the LIM map with signal and all contaminants included averaged over 1,000 realizations, with the purple shaded region around it being the standard deviation of the realizations. The blue dashed curve is for a LIM map with just [CII] signal and the red dashed curve is for a LIM map with just CO interlopers and Galactic foregrounds. The instrument resolution convolves all maps used. (a) exhibits significant contamination at low $k$ due to foregrounds. (b) depicts the same spectra but with the lowest $k_x$, $k_y$, and $k_z$ modes removed.}
    \label{fig:cross_pow}
\end{figure*}

Fig. \ref{fig:cross_pow}(a) depicts the cross power spectra calculated from a set of LIM maps and mock QSO catalogs. At most $k$, the cross power recovers the [CII] cross power spectrum well. At low and high $k$, the noise is considerable due to the low number of modes and low signal-to-noise, respectively, leaving a middling $k$ range in which useful information can be extracted. However, at low and middling $k$, we find that the Galactic foregrounds have a noticeable contribution to the cross power.\footnote{While Fig. \ref{fig:cross_pow} shows the cross power of the CO interlopers and foregrounds combined, removing the foregrounds from the maps eliminates the contamination, so we can isolate the effect to the foregrounds alone.} Contamination was also observed in the ratio of the Fourier-transformed CVIDs, despite the mathematical independence of such fields (see Sec. \ref{sec:cvid}). The source of this spurious correlation is not entirely clear, as our foreground model (see Sec. \ref{sec:foregrounds}) is generated independently of the halo catalogs. Along the frequency dimension, the correlation might arise from both the foreground brightness and the QSO density increasing simultaneously, but the true source of the correlation here is not well understood. We leave the determination of this source and how to recover the low-$k$ modes from it to future work. Generally, low-$k$ modes suffer from numerous problems, like the low number of modes, non-Gaussian errors, and large-scale effects like 1/$f$ noise found in real data (but not present in our simulations), so such modes might be removed or ignored in an analysis regardless.

To remedy this contamination, we institute both a map-space and a $\textbf{k}$-space filter, as used in similar analyses (see, e.g., \citet[]{kerrigan_improved_2018}). In map-space, we remove from the analysis regions of space (corresponding to whole $3\degree\text{ R.A.}\times2.5\degree\text{ Dec.}$  patches) where the mean foreground intensity is larger than $\sim1$ MJy/sr. This corresponds to the portion of S82 from $-45\degree<\text{R.A.}<-39\degree$. Any QSOs identified within this region are omitted from calculations. In $\textbf{k}$-space, we conservatively remove only the $0$ Mpc$^{-1}$ mode of $k_x$, $k_y$, and $k_z$ (R.A., Dec., and frequency/redshift, respectively, in comoving coordinates) from the maps. We apply these filters to another set of simulated LIM maps prior to the cross-correlation analyses, and all results discussed hereafter, including those of stacking and the CVID, use these filtered simulations.

As seen in Fig. \ref{fig:cross_pow}(b), these filters reduce the contribution of the foregrounds considerably while preserving most of the [CII] signal. However, any positive power in the lowest $k$ bin is lost. When computing SNRs and constraining $A$, we omit this bin. The use of these filters, along with the use of noise weighting could introduce covariance across $k$-space. By running many simulations, we model that information and can see that such covariances are minimal (see Fig. \ref{fig:corr}).\footnote{The variance in the total LIM cross power comes almost entirely from the noise and the CO/foregrounds. The variance of the [CII]-only cross power is so minimal that it would not be visible if plotted in Fig. \ref{fig:cross_pow}.}

Using Eq. \ref{eq:cross_snr}, we compute the mean SNR of the cross power spectrum to be 8.42, with SNRs of 6.89 and 9.99 at the 2.5 and 97.5 percentiles, respectively. Even with our conservative approach to foreground filtering, the additional spatial information available to the cross power spectrum compared to the other two cross-correlation techniques improves the detectability of [CII] (see Table \ref{tab:SNRs}).

\subsection{Stacking Results}

\begin{figure*}
    \centering
    \begin{subfigure}[t]{0.33\textwidth}
        \centering
        \includegraphics[width=\textwidth]{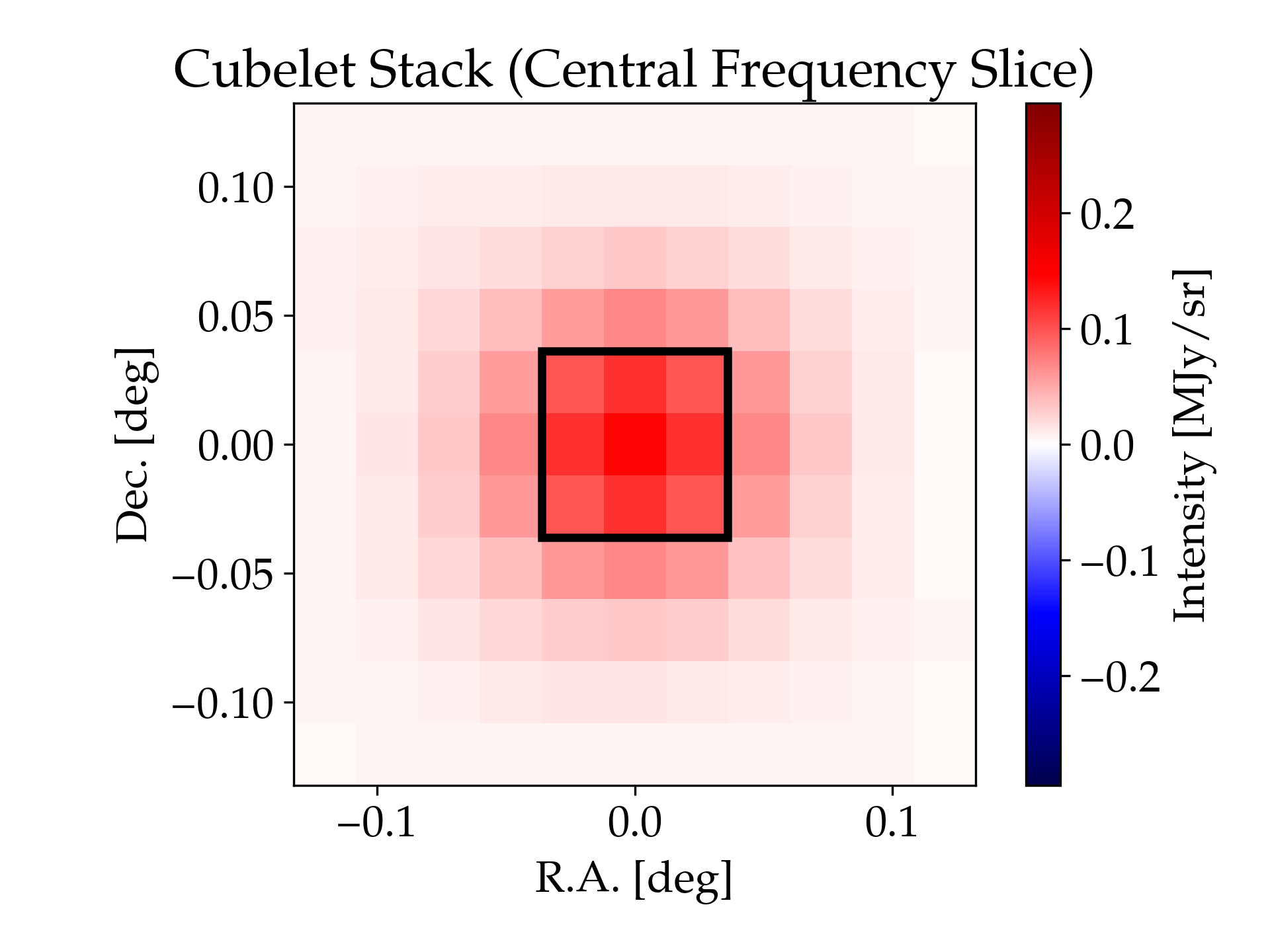}
        \caption{QSO-centered, [CII] only}
    \end{subfigure}%
    ~ 
    \begin{subfigure}[t]{0.33\textwidth}
        \centering
        \includegraphics[width=\textwidth]{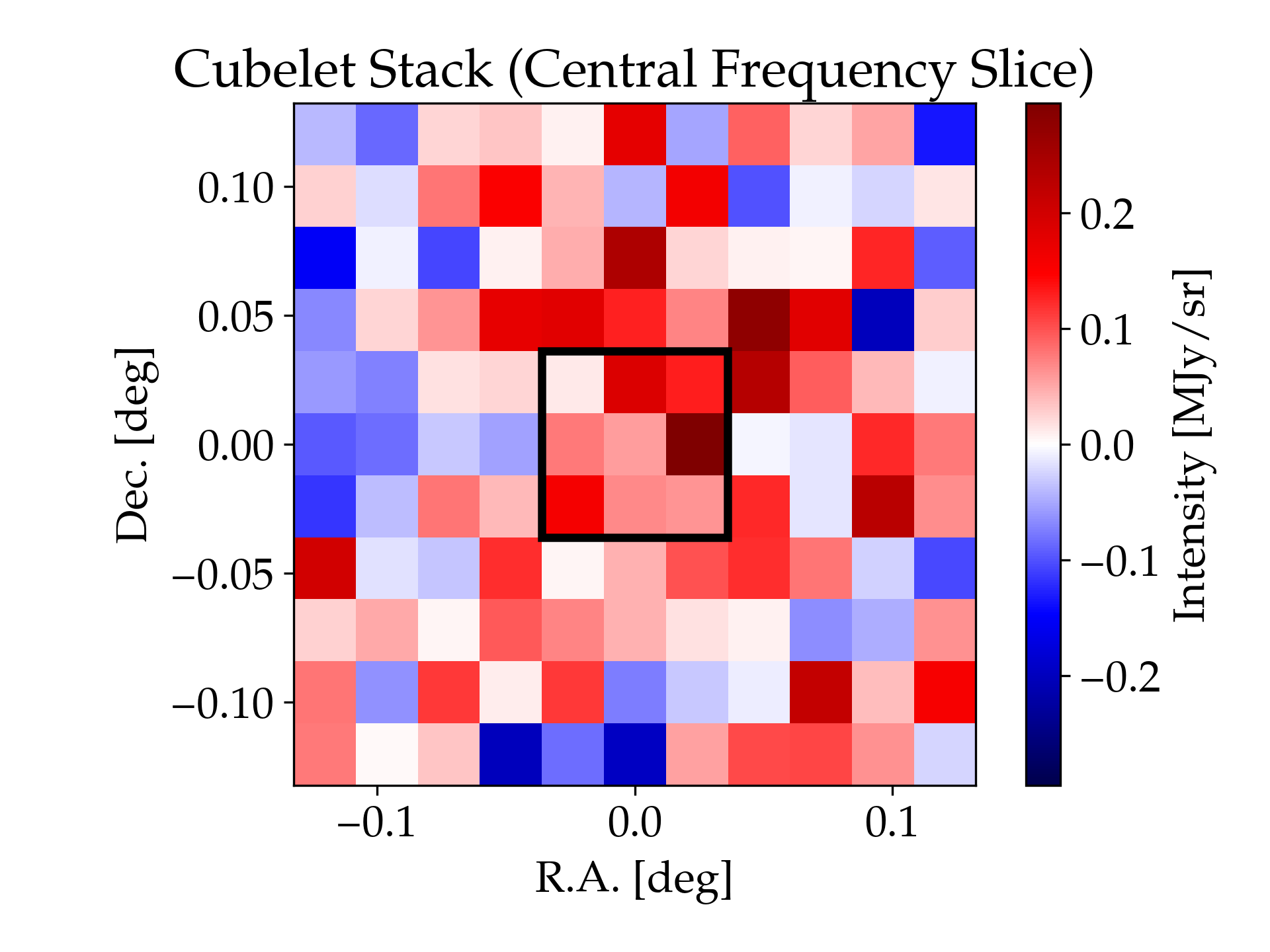}
        \caption{QSO-centered, full map with noise}
    \end{subfigure}%
    ~ 
    \begin{subfigure}[t]{0.33\textwidth}
        \centering
        \includegraphics[width=\textwidth]{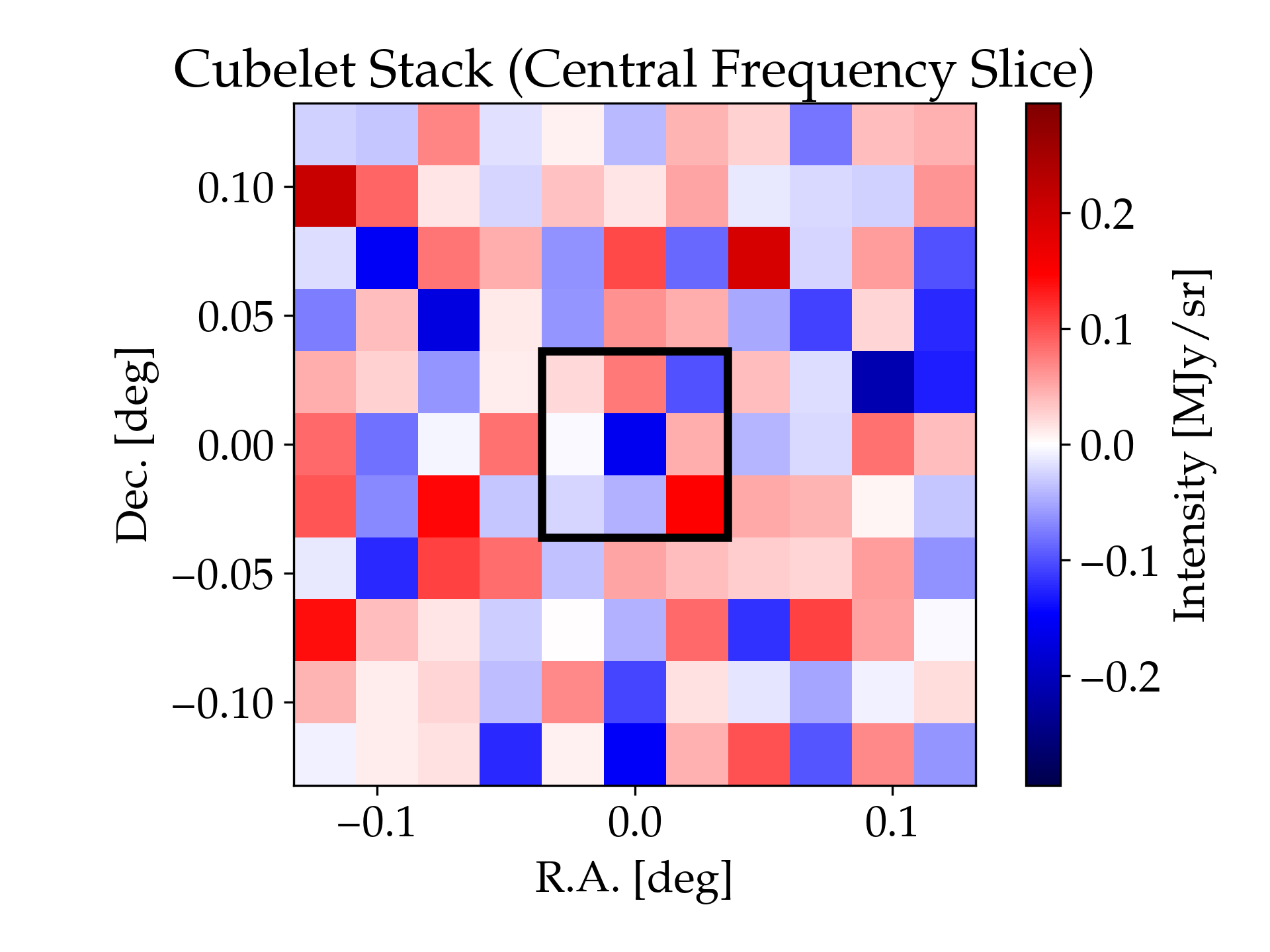}
        \caption{Random locations, full map with noise}
    \end{subfigure}
    \caption{Example stacks resulting from the stacking process. These depict the frequency slice of the stack in which the QSO (or random location) is located. Coordinates are relative to the center. All stacks are scaled according to the minimum/maximum values of the stack in (b). The black square outlines the central 9 voxels of the stack slice that will be included in the summed intensity observable for the stack.}
    \label{fig:ang_stacks}
\end{figure*}

\begin{figure*}
    \centering
    \begin{subfigure}[t]{0.33\textwidth}
        \centering
        \includegraphics[width=\textwidth]{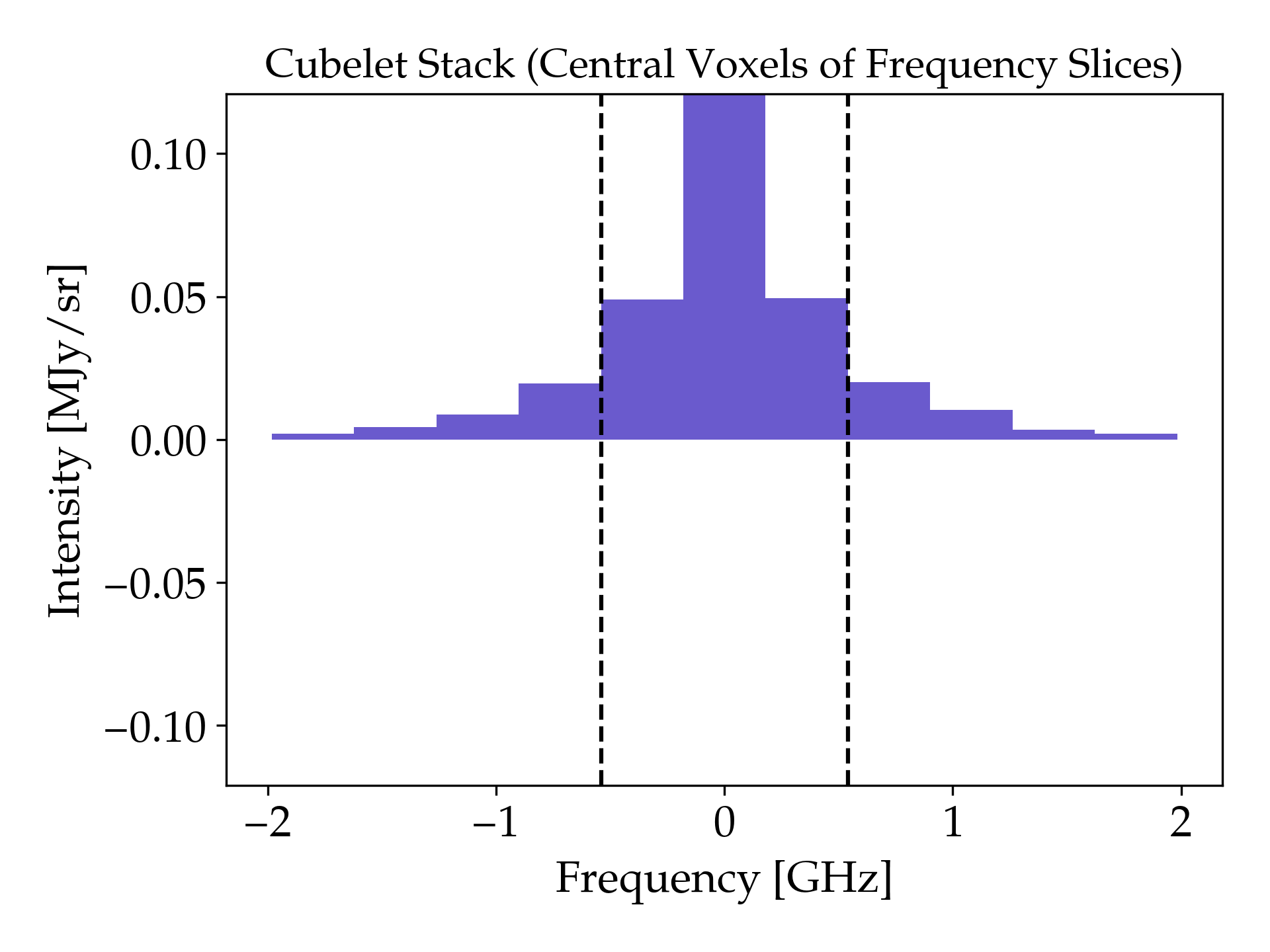}
        \caption{QSO-centered, [CII] only}
    \end{subfigure}%
    ~ 
    \begin{subfigure}[t]{0.33\textwidth}
        \centering
        \includegraphics[width=\textwidth]{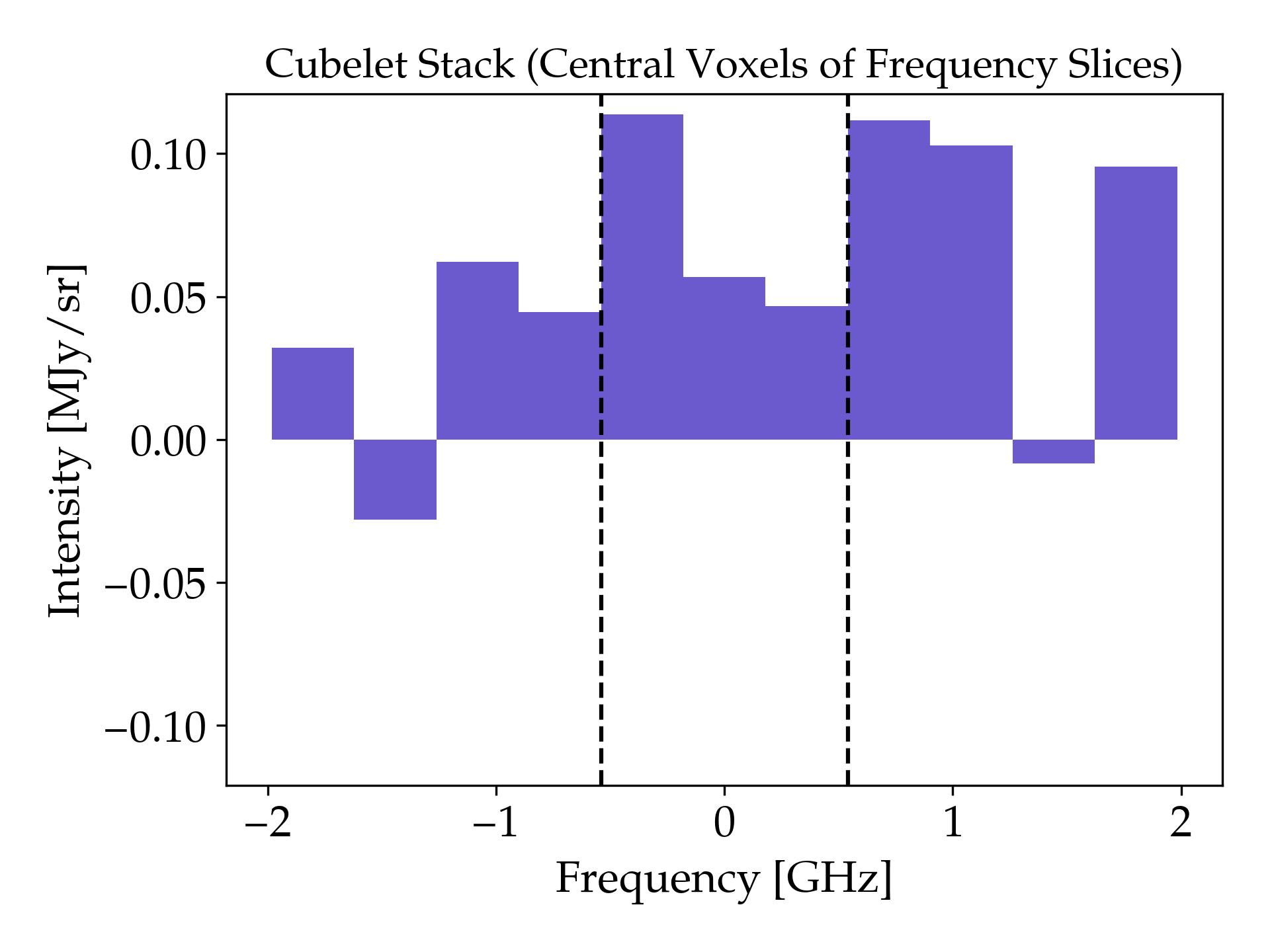}
        \caption{QSO-centered, full map with noise}
    \end{subfigure}%
    ~ 
    \begin{subfigure}[t]{0.33\textwidth}
        \centering
        \includegraphics[width=\textwidth]{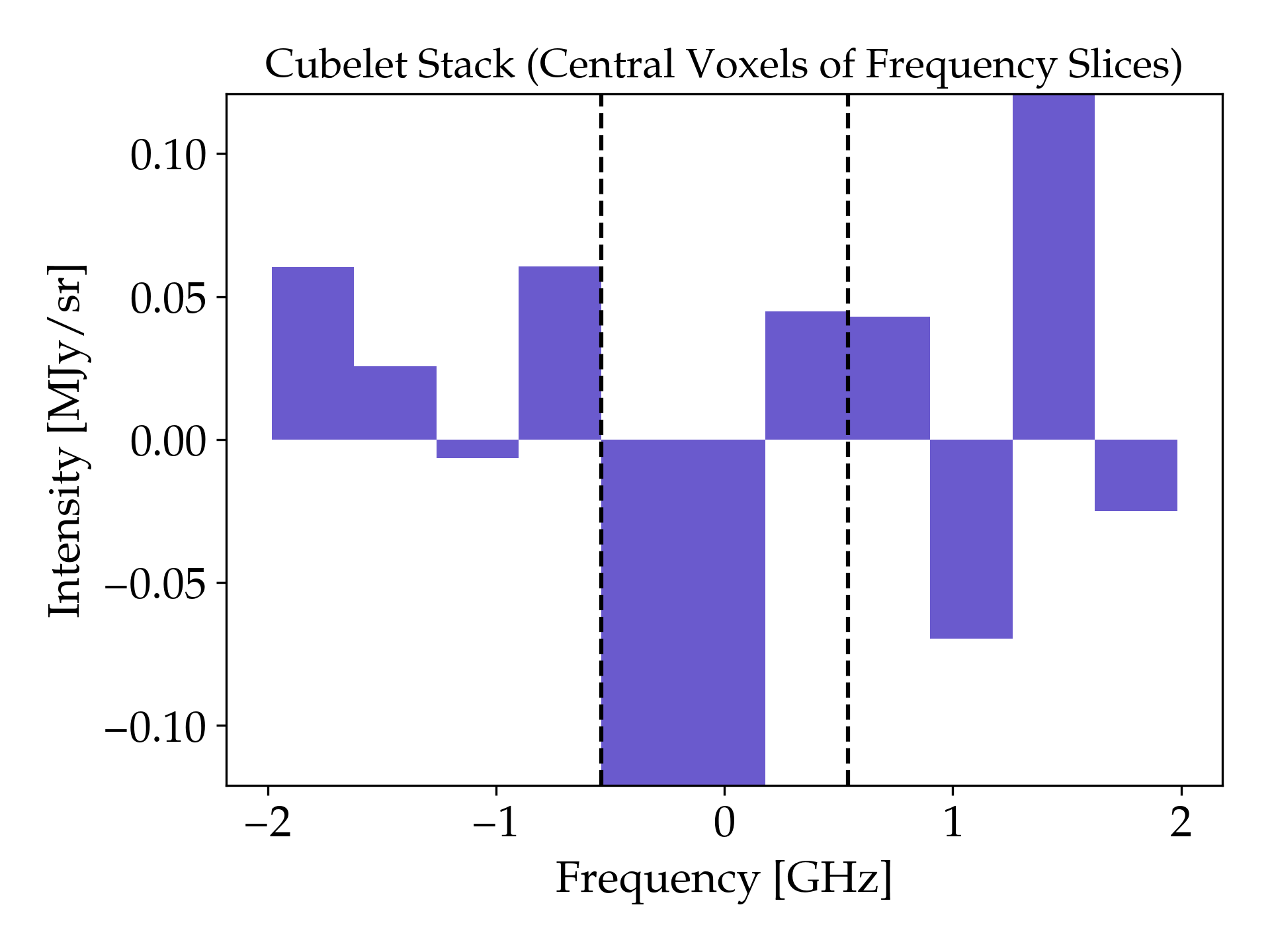}
        \caption{Random locations, full map with noise}
    \end{subfigure}
    \caption{Example stacks resulting from the stacking process. These depict the central voxels of the stack across frequency slices. Coordinates are relative to the center. All stacks are scaled according to the minimum/maximum values of the stack in (b). The black dashed lines outline the central 3 voxels of the stack  that will be included in the summed intensity observable for the stack.}
    \label{fig:freq_stacks}
\end{figure*}

Examples of stacks performed on the LIM maps can be found in Fig. \ref{fig:ang_stacks} and Fig. \ref{fig:freq_stacks}. Figs. \ref{fig:ang_stacks}(a) and \ref{fig:freq_stacks}(a) illustrate a single stack performed on a LIM map without any instrument noise. Most of the intensity present is concentrated near the center and is consistent with a bright source in the central voxel convolved by the EXCLAIM beam. However, as was shown in \citet[]{dunne_three-dimensional_2025} and confirmed in our own testing, nearby halos clustered with the QSO contribute  considerable ($\sim$50\%) intensity to the total observed.

\begin{figure}
    \centering
    \includegraphics[width=\linewidth]{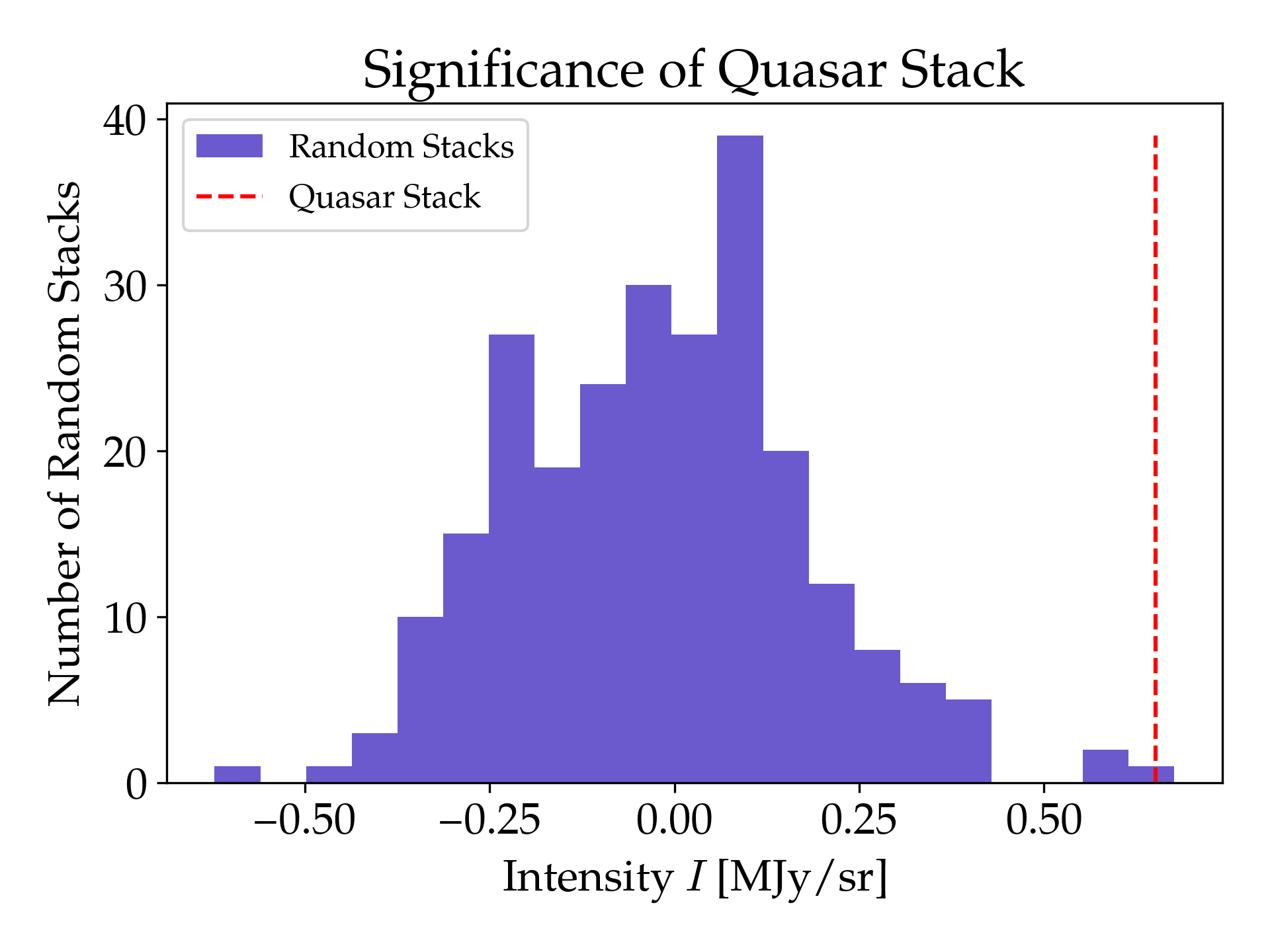}
    \caption{An example of a stacking bootstrap test. The red dashed line indicates the weighted sum of the central 27 voxels of the stack depicted in Figs. \ref{fig:ang_stacks}(b) and \ref{fig:freq_stacks}(b). The violet histogram is the distribution of the weighted sums for 250 stacks performed on random locations on the LIM map.}
    \label{fig:sig_test}
\end{figure}

In Figs. \ref{fig:ang_stacks}(b) and \ref{fig:freq_stacks}(b), noise has been added to the stacks in their respective (a)s. It is difficult to distinguish the signal visually in these plots. Figs. \ref{fig:ang_stacks}(c) and \ref{fig:freq_stacks}(c) depict an example of a stack from the same noise instance as their respective (b)s but performed on random locations in the map. In these single examples, it is not always clear the QSO-centered stack is distinguishable from the random location baseline. When the QSO-centered stack is compared to a distribution of such random stacks, as outlined in Sec. \ref{sec:stacking}, it becomes more apparent our example stack might be brighter than a typical random location-centered stack (see Fig. \ref{fig:sig_test}).

Performing the QSO-centered stacking process 1,000 times, each with 250 random location-centered stacks, we compute 1,000 SNRs using Eq. \ref{eq:stacking_snr}. The mean stacking SNR is 4.47, with the 2.5 and 97.5 percentile SNRs being 2.51 and 6.63, respectively.

\subsection{CVID Results}\label{sec:cvidresults}

\begin{figure*}
    \centering
    \begin{subfigure}[t]{0.49\textwidth}
        \centering
        \includegraphics[width=\textwidth]{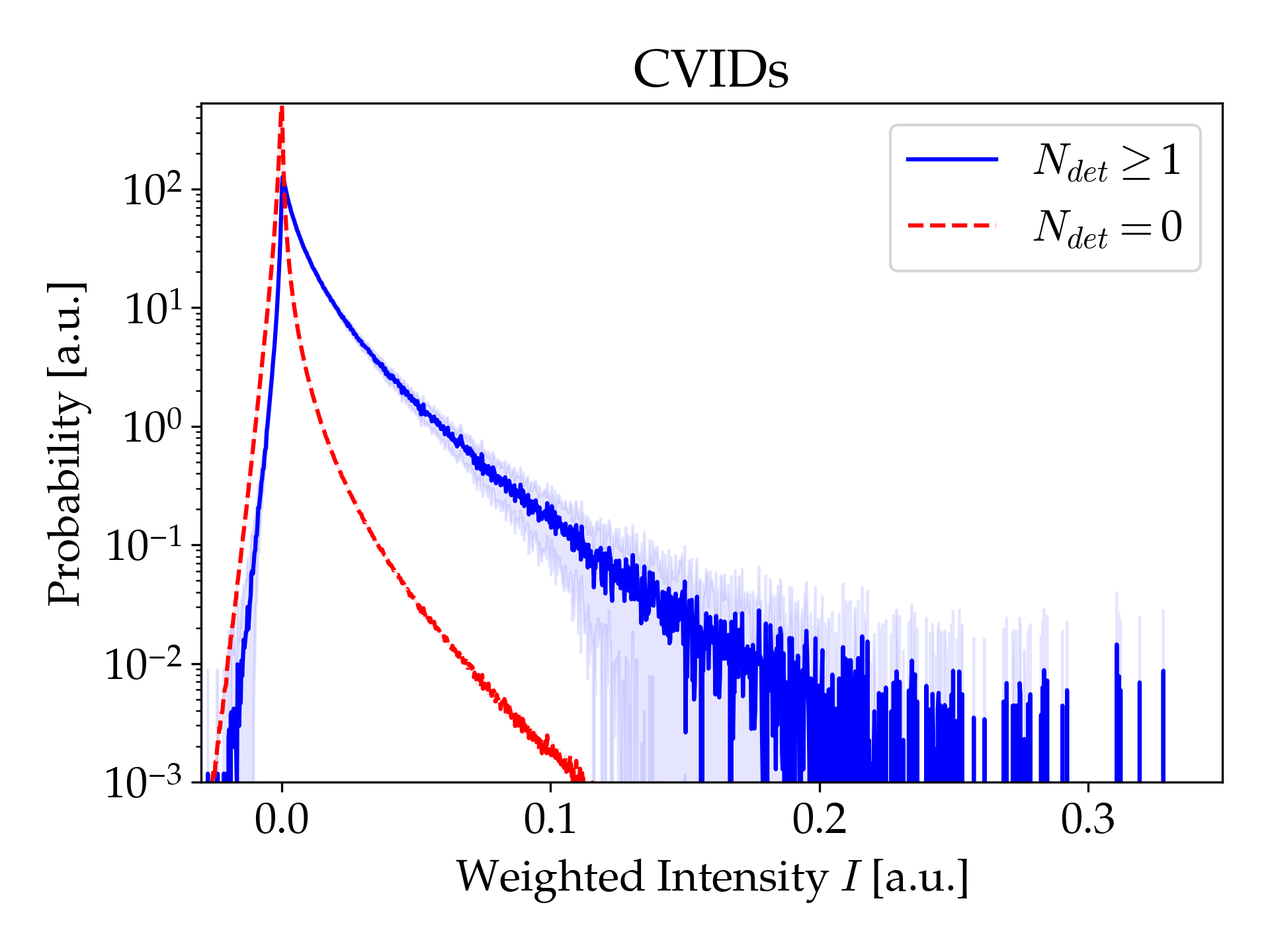}
        \caption{[CII] Only}
    \end{subfigure}%
    ~ 
    \begin{subfigure}[t]{0.49\textwidth}
        \centering
        \includegraphics[width=\textwidth]{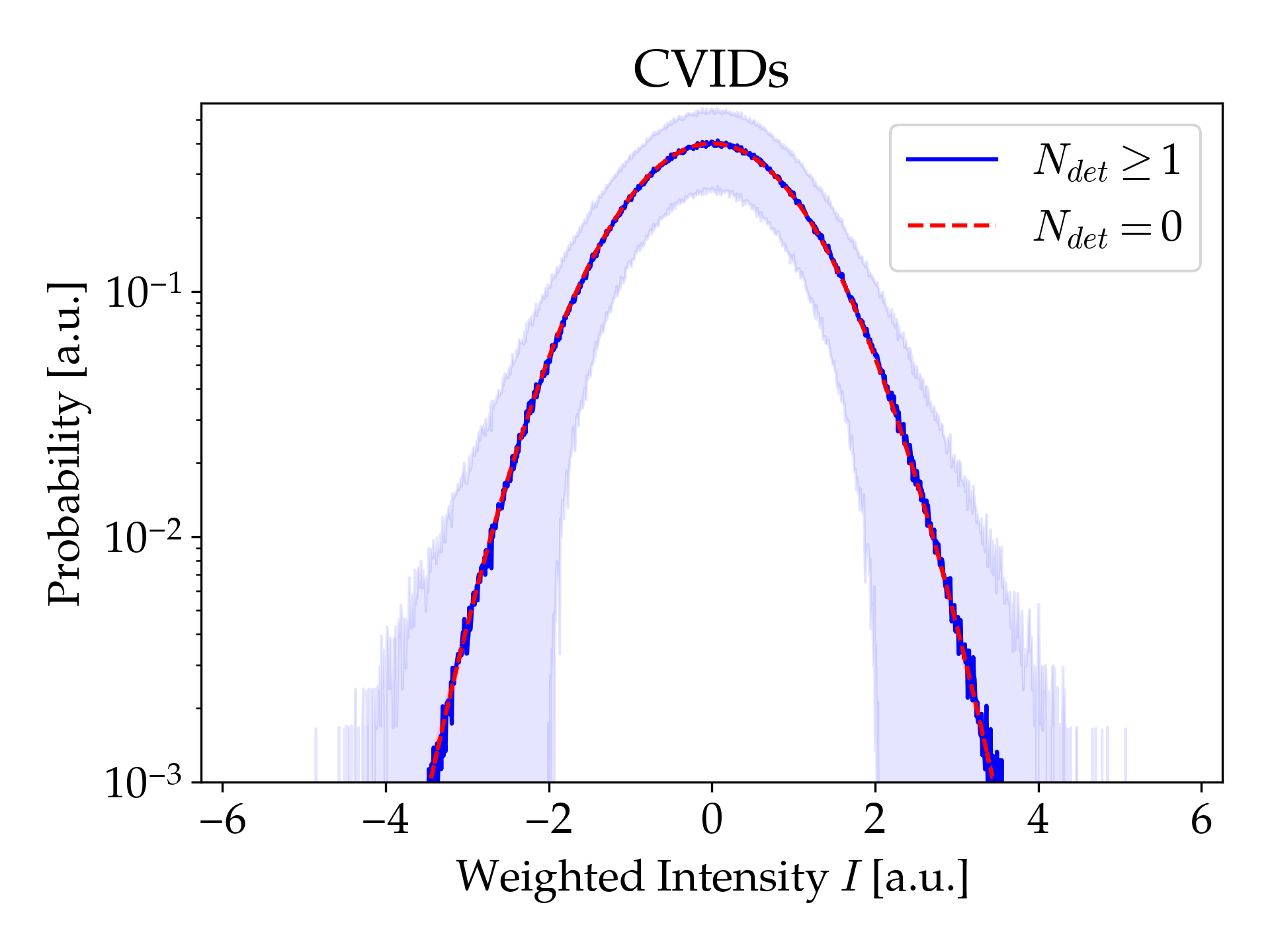}
        \caption{Full map with noise}
    \end{subfigure}
    \caption{The mean CVIDs produced from the LIM maps. The solid blue curve is the mean CVID for voxels containing a QSO and the immediately adjacent voxels. The dashed red curve is the mean CVID for all other voxels. Each curve is accompanied by a shaded region of the same color showing the one standard deviation error of each bin. Note that the $N_{\text{det}}=0$ curve's errors are too small to be seen. In (b), bins are omitted at regular intervals for ease of visualization. This omission is not utilized in any calculations.}
    \label{fig:cvids}
\end{figure*}

The mean CVIDs are shown in Fig. \ref{fig:cvids}. Fig \ref{fig:cvids}(a) is a CVID created with the LIM map containing no noise. The $N_{\text{det}}\geq1$ curve (blue) shows a greater probability at higher weighted intensities than the $N_{\text{det}}=0$ curve (red). This is suggestive of the higher intensity emission from the QSO-containing halos, but the effect is visibly washed out when noise is added to the map, as seen in Fig. \ref{fig:cvids}(b). Here, the strong, Gaussian noise results in visually identical CVIDs, regardless of the presence of a QSO, demonstrating the necessity of an estimator like Eq. \ref{eq:ratio}.

\begin{figure*}
    \centering
    \begin{subfigure}[t]{0.49\textwidth}
        \centering
        \includegraphics[width=\textwidth]{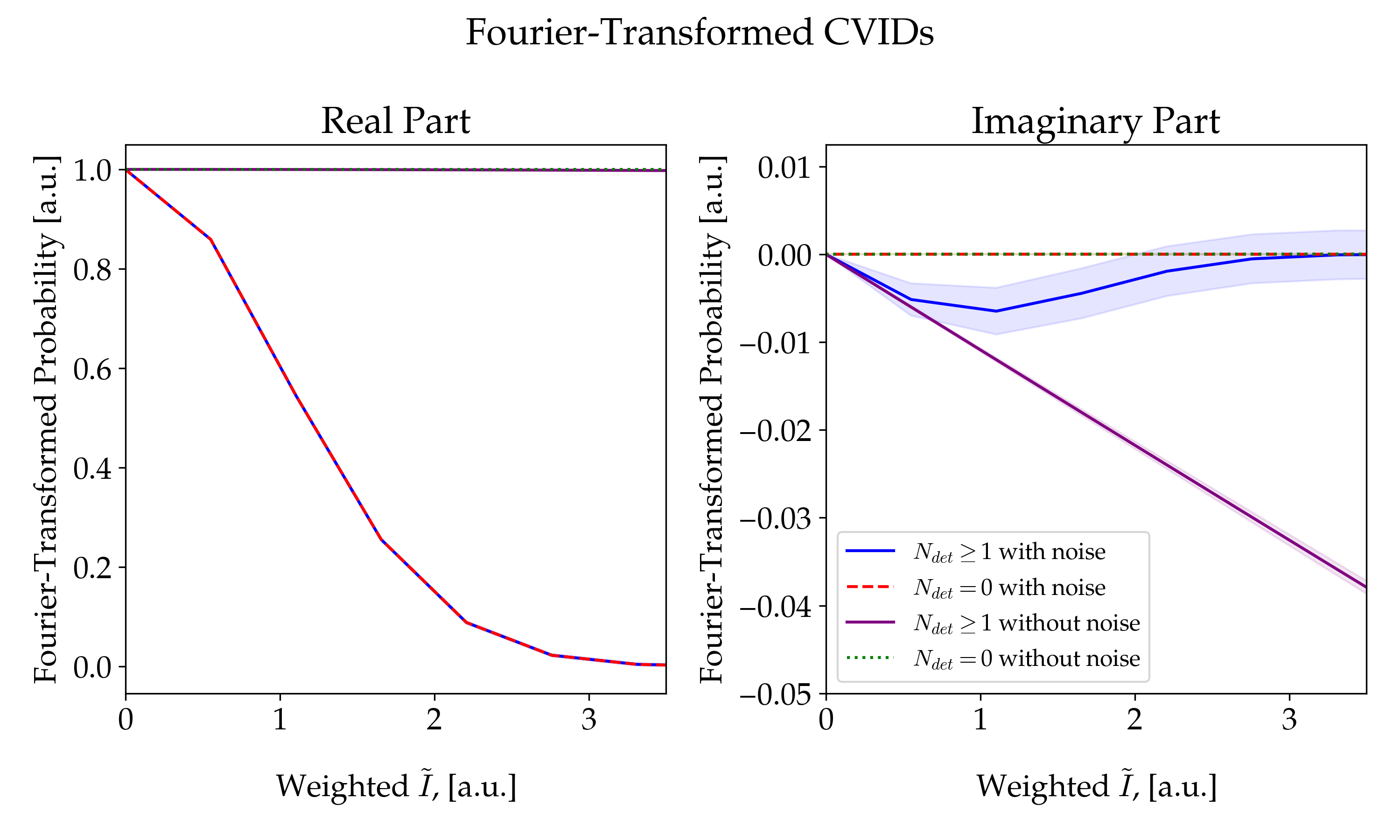}
        \caption{Full noise}
    \end{subfigure}%
    ~ 
    \begin{subfigure}[t]{0.49\textwidth}
        \centering
        \includegraphics[width=\textwidth]{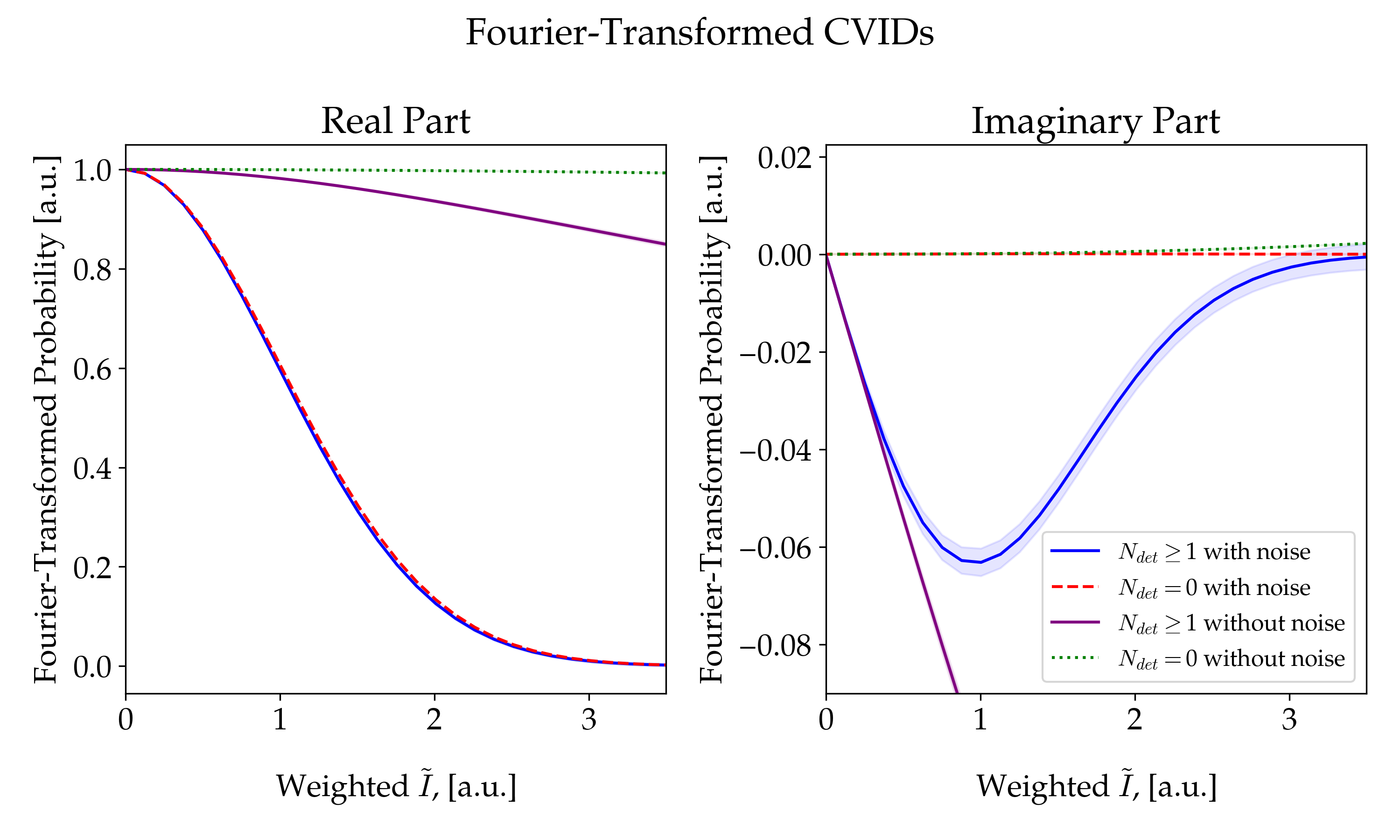}
        \caption{Reduced noise}
    \end{subfigure}
    \caption{The mean Fourier-transformed CVIDs. The solid blue curves and dashed red curves were calculated from the CVIDs created from LIM maps with noise, while the solid navy curves and dashed maroon curves correspond to LIM maps without noise. Shaded one standard deviation error regions are also displayed, but are largely too small to see. The Fourier-transformed CVIDs in (a) come from maps with the full EXCLAIM-like noise levels, while those in (b) come from maps with a reduced ($\times0.1$ NEI) noise level. Note the difference in vertical scaling between the subplots.}
    \label{fig:ft_cvids}
\end{figure*}

In Fig. \ref{fig:ft_cvids}(a), we plot the Fourier-transformed CVIDs as calculated in Eq. \ref{eq:ft_cvid}. Visually, it is still difficult to distinguish between the $N_{\text{det}}\geq1$ (blue) and $N_{\text{det}}=0$ (red) cases when noise is considered. The Fourier-transformed CVIDs of the noiseless cases (navy and maroon, respectively), are considerably different from the noisy counterparts and distinguishable from each other. Note that we only plot the positive weighted $\tilde{I}$ domain as the Fourier-transformed CVIDs are symmetric across the $y$-axis; it is an even function for the real part and an odd function for the imaginary part. Given this, we consider only the positive weighted $\tilde{I}$ bins in the remainder of the analysis as no new information is gained from the negative bins.

\begin{figure*}
    \centering
    \begin{subfigure}[t]{0.49\textwidth}
        \centering
        \includegraphics[width=\textwidth]{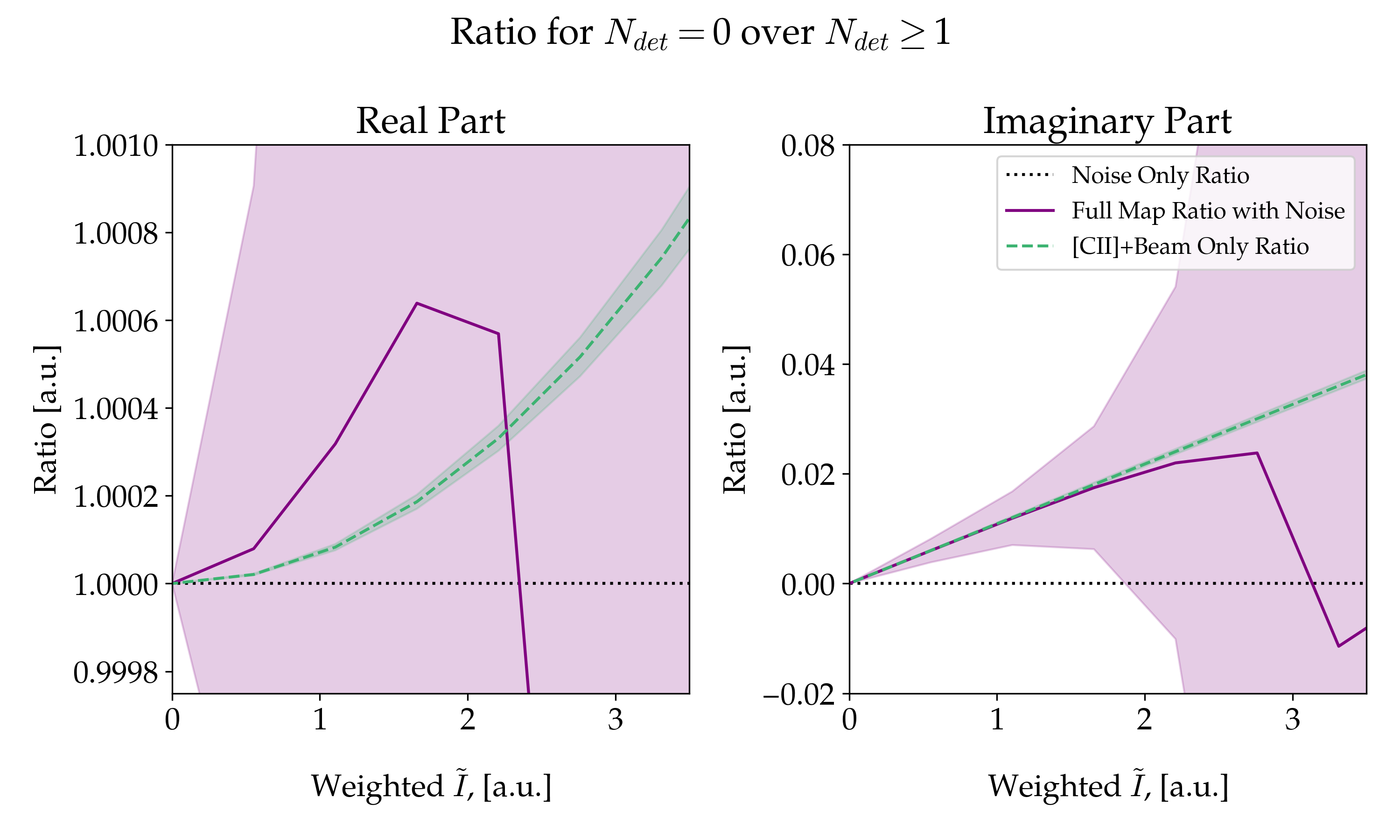}
        \caption{Full noise}
    \end{subfigure}%
    ~ 
    \begin{subfigure}[t]{0.49\textwidth}
        \centering
        \includegraphics[width=\textwidth]{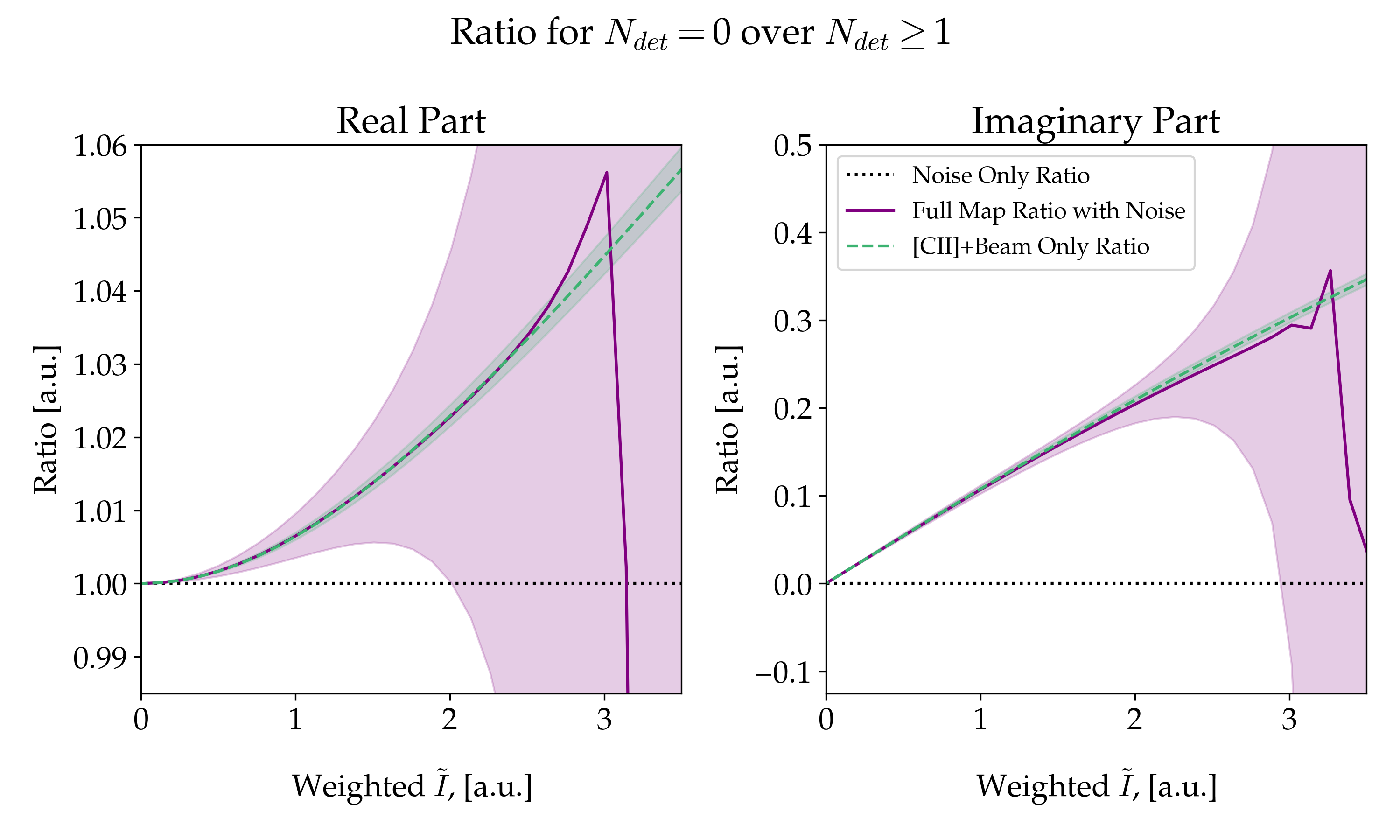}
        \caption{Reduced noise}
    \end{subfigure}
    \caption{The mean ratio of the Fourier-transformed CVIDs. The solid purple curves correspond to the observable calculated for LIM maps with noise, while the dashed green curves correspond to LIM maps without noise. Shaded one standard deviation error regions of the same color are also displayed. The ratio of Fourier-transformed CVIDs in (a) come from maps with the full EXCLAIM-like noise levels, while those in (b) come from maps with a reduced ($\times0.1$ NEI) noise level. Note the difference in vertical scaling between the subplots.}
    \label{fig:cvid_ratios}
\end{figure*}

In Fig. \ref{fig:cvid_ratios}(a), we plot the CVID ratio from Eq. \ref{eq:ratio}. The utility of this observable in removing the impact of uncorrelated fields is demonstrated in that the ratio computed from LIM maps with noise (purple) is largely identical to that computed without noise (green). Notably, however, there is a certain weighted $\tilde{I}$ at which the ratio with noise begins to diverge from its noiseless counterpart and oscillates around 0. At sufficiently high signal-to-noise ratios, this is the expected behavior from a ratio of two, independent, Gaussian-distributed variables. For an in-depth explanation, see Appendix \ref{ap:patrick}. Including weighted $\tilde{I}$ beyond this point can greatly skew SNR and $A$ constraint calculations by spuriously showing a significant deviation from the noise-only equivalent.

\begin{figure*}
    \centering
    \begin{subfigure}[t]{0.49\textwidth}
        \centering
        \includegraphics[width=\textwidth]{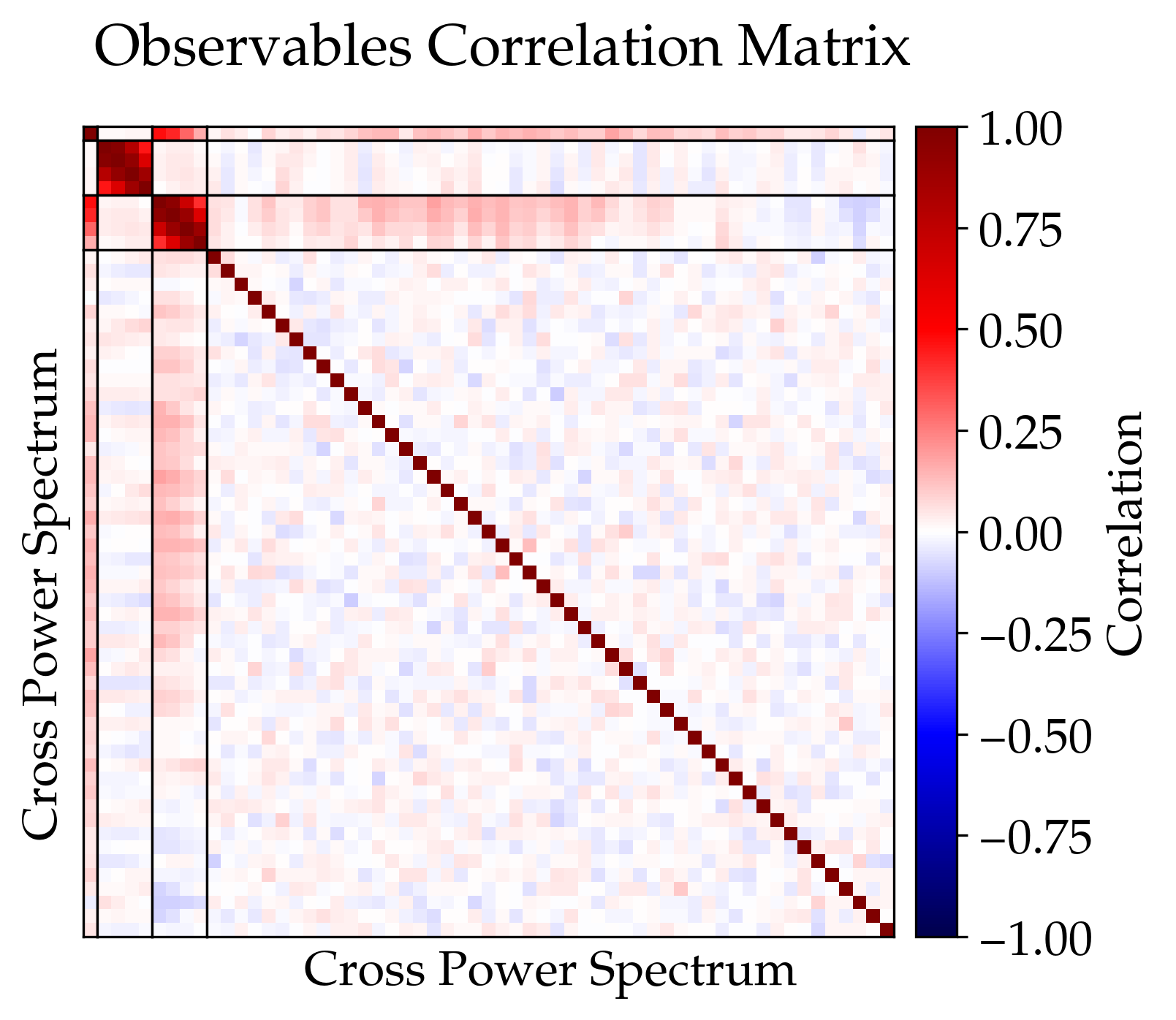}
        \caption{Full Correlation}
    \end{subfigure}%
    ~ 
    \begin{subfigure}[t]{0.49\textwidth}
        \centering
        \includegraphics[width=\textwidth]{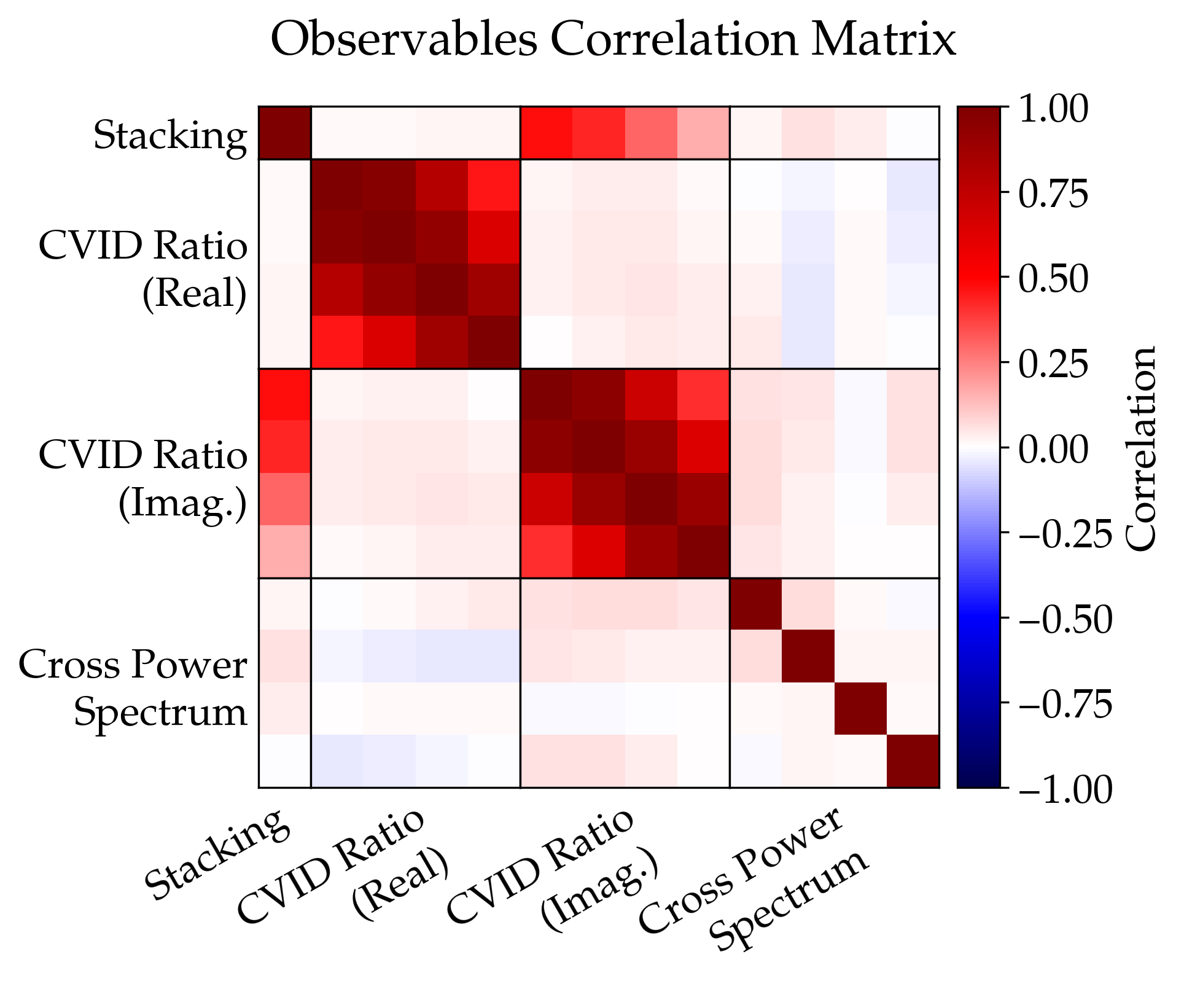}
        \caption{Upper Left Corner}
    \end{subfigure}
    \caption{The correlation matrix of the three observables as described in Sec. \ref{sec:disc}. Each pixel represents one bin of the observable. (a) is the full correlation matrix, while (b) is the upper left corner of (a) enlarged for visibility. The correlation matrix is unitless and can range from $+1$ (directly correlated) to $-1$ (inversely correlated).}
    \label{fig:corr}
\end{figure*}

To accommodate this behavior, we institute a cut-off in weighted $\tilde{I}$ at 2.5. Using Eq. \ref{eq:cvidsnr}, we compute a distribution of SNRs for the ratio observable. Note that most of the information of the ratio SNR comes from the imaginary part, which diverges more significantly from 0. Information from the correlation between bins is also not negligible, as can be seen in the CVID-CVID portion of Fig. \ref{fig:corr}. The mean SNR of the ratio is 3.92, with SNRS of 2.32 and 5.69 at the 2.5 and 97.5 percentiles, respectively. 

A CVID and stack should capture very similar, if not identical information. Consider a [CII] signal-only LIM map with no beam effects or large-scale clustering. A QSO-centered stack's central voxel's intensity should (and does) match the first moment of the $N_{\text{det}}\geq1$ CVID. In this idealized case, the stack would contain purely a subset of the CVID information, as the CVID incorporates all of the other moments of the $N_{\text{det}}\geq1$ distribution as well as the much larger number of $N_{\text{det}}=0$ voxels. However, there are several subtleties which modify the easy comparison between the two in our simulations. First of all, the ratio step in the CVID analysis sacrifices some hard-to-define amount of statistical information in the name of systematics rejection. Second, beam and large-scale structure effects distribute emission from the central voxel into its surroundings, which is handled in different ways by the two estimators. Finally, our simple, single-parameter model removes some of the comparative advantage of the CVID over the stack. Since stacking as we have done here compresses the map to a single average measurement, parameter inference will always suffer from degeneracies if more than one parameter is included. The CVID, on the other hand, has multiple $\mathcal{R}$ bins to select from and can at least in principle constrain more complex models.  We leave further exploration of the relative importance of these factors for future work.

\subsection{Constraining Intensity Model Results}\label{sec:constrainAresults}


With these simulations, we attempt to constrain the $A$ parameter as detailed in Sec. \ref{sec:constrainA}. For the CVID ratio, we use the first four, positive weighted $\tilde{I}$ bins as explained in Sec. \ref{sec:cvidresults}, and for the cross power spectrum, we omit the lowest $k$ bin as explained in Sec. \ref{sec:crosspowerresults}. For constraining $A$ with one or a combination of observables, we use only the portion of the covariance matrix corresponding to the appropriate observables.

\begin{figure}
    \centering
    \includegraphics[width=\linewidth]{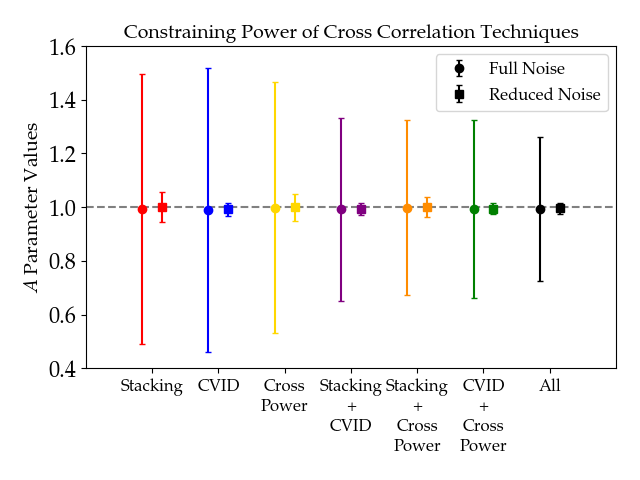}
    \caption{The constraining power of the cross-correlation techniques for a fiducial value of $A=1$ (dashed, black line). Circles/squares represent the most probable value for $A$ for the full/reduced noise levels, respectively, while the error bar bounds are the 95\% confidence interval lower and upper bounds, as listed in Table \ref{tab:A constraints}.}
    \label{fig:constrainA}
\end{figure}

As illustrated in Fig. \ref{fig:constrainA} and listed in Table \ref{tab:A constraints}, the constraining power of any cross correlation technique by itself for the fiducial $A=1$ model is approximately 50\%. When two techniques are used together, constraining power is increased to within 33\% of the fiducial value. All three techniques used together constrain at about 27\%. Under the assumption that the distribution for $A$ is Gaussian, we report the number of standard deviations between $A=0$ and the most probable value of $A$. Under this framing, we can reject $A=0$ at $\sim4\sigma$ using just one technique, at $\sim6\sigma$ for two techniques together, and at $\sim7.3\sigma$ for all three together.

\subsection{Reduced Noise Scenario}\label{sec:reducedresults}

While our results suggest a first-generation LIM experiment like EXCLAIM would be capable of detecting its target line and constraining an intensity model, its ability to do so would be limited. This is in line with forecasts done for other first generation LIM experiments, as in, e.g., \citet[]{chung_comap_2022}, \citet[]{karkare_spt-slim_2022}, and \citet{ccat-prime_collaboration_ccat-prime_2023}. We therefore consider a more futuristic EXCLAIM-like instrument in which the noise level has been reduced to 10\% of its current level. This is a reasonable improvement for an instrument with about 100 times more detectors, or a mission that has 100 times more observing time. Alternatively, one could envision a space-based mission with no atmospheric foreground whatsoever. We perform the simulations and analysis pipelines in a nearly (see below) identical manner as explained above, and the results are summarized in Tables \ref{tab:SNRs} and \ref{tab:A constraints}.

As expected, the SNRs one can achieve with any of the techniques increase substantially. The mean SNRs from our simulations are 44.01, 24.62, and 34.26 for stacking, the CVID ratio, and the cross power spectrum, respectively. Similarly, the three cross-correlation techniques constrain $A$ to a much greater extent. The constraining power is about 5\%, 3\%, and 2\% for one, two, and three techniques simultaneously, respectively, though there is some slight variation from technique to technique. Alternatively, we can say that we reject $A=0$ at $\sim35-76\sigma$, $\sim52-95\sigma$, and $\sim96\sigma$ for one, two, and three techniques simultaneously, respectively. Figures equivalent to \ref{fig:cross_pow}(b), \ref{fig:ang_stacks}, \ref{fig:freq_stacks}, \ref{fig:cvids}, and \ref{fig:corr} for this low noise scenario can be found in Appendix \ref{ap:lownoise}.

In Figs. \ref{fig:ft_cvids}(b) and \ref{fig:cvid_ratios}(b), we show the Fourier-transformed CVIDs and their ratio in the reduced noise scenario to show how the procedure changes with noise level. Note that we choose different bin widths in the two cases. Since "binning" is a purely numerical operation that should have no effect on the final constraints, we select the optimal bin size for each noise level by ensuring that they produce a converged likelihood distribution on $A$. We consider the likelihood to be adequately converged when the maximum probability $A$ is within 1\% of the true value with a similar degree of distribution symmetry.


In theory, the Fourier-transformed, noiseless $\ndet{0}$ and $\ndet{1}$ CVIDs would be identical in Figs. \ref{fig:ft_cvids}(a) and (b). However, due to the change in noise level, we observe slight differences in these two curves. Similarly, in Figs. \ref{fig:cvid_ratios}(a) and (b), all of the depicted ratios should be identical across noise levels, but are not. This is due to the difference in voxel weightings. When the noise level (NEI) is reduced, the weights are reduced by the same amount, which can alter the shape of the curves. Even so, the ratios for both the original and reduced noise levels display the behavior in which the real and imaginary parts descend to 0 around the threshold of weighted $\tilde{I}=2.5$.

\section{Discussion}\label{sec:disc}

Under the fiducial \citet[]{padmanabhan_constraining_2019} model and current EXCLAIM instrument noise levels, we find that an EXCLAIM-like instrument could reasonably make a strong detection of the cosmological [CII] signal using cross-correlation techniques. The cross power spectrum provides the best opportunity for such a detection, as it reliably provides the highest SNRs. Stacking and the CVID could also offer opportunities for detection, with a slight preference for stacking.

The three techniques appear to capture different sets of information. To illustrate this, consider the correlation matrix $\textbf{R}$ of the observables, calculated using the covariance matrix $\textbf{C}$ according to
\begin{equation}
    \textbf{R} = \text{Diag}(\textbf{C})^{-\frac{1}{2}}\:\textbf{C}\:\text{Diag}(\textbf{C})^{-\frac{1}{2}}.
\end{equation}
The correlation matrix $\textbf{R}$ is depicted in Fig. \ref{fig:corr}. If the correlation between two observable bins is positive, they are directly correlated, with strength of correlation increasing as the correlation approaches $+1$. Similarly, if the correlation is negative, they are inversely correlated with a strength increasing as the correlation approaches $-1$.

The correlation matrix reveals a few points of interest. First, the CVID ratio bins are highly correlated with each other as expected, but there is no strong correlation between the real and imaginary parts. Second, the CVID ratio and stacking techniques have the highest cross-technique correlation. Notably, this correlation only exists between the imaginary part of the ratio and stacking. This suggests that the CVID and stacking techniques capture some mutual information, but still have some unique contributions to the overall picture. Finally, the cross power spectrum appears to have a low-to-moderate positive correlation at middling $k$ bins with both stacking and imaginary part of the CVID ratio. One can say that, while the cross power is partially redundant with respect to the one-point statistics, it contains the most unique set of information relative to other techniques.

Despite the three techniques measuring different sets of information about the [CII] signal, their abilities to constrain $A$ are nearly identical. The cross power spectrum appears to do marginally better than stacking and the CVID, but the small difference is an interesting contrast to the significantly better performance of the cross power spectrum when it comes to SNRs. 

We note that, when comparing the mean SNRs of the individual techniques to their corresponding individual constraining power significances, there is a slight difference for the stacking and CVID techniques, and a more considerable difference for the cross power spectrum. In the low noise scenario, the stacking and cross power techniques become more comparable across the two analyses, while the CVID takes on a much larger difference. We ascribe the difference between the SNR and $A$ constraint calculations primarily to the fact that they incorporate the information contained within the LIM maps in different ways. The probability distribution of $A$ is not necessarily Gaussian, resulting in confidence intervals which do not match the shape of the SNR distributions. Additionally, in the particular case of the CVID, the dependency on $A$ is not simple. Given the Fourier transforms used in the statistic, $A$ can serve to phase shift signal, changing its interaction with the convolution. By changing the noise level, we also change which techniques match better across calculations, indicating that the techniques respond differently to varying noise levels. We leave a formal diagnosis of these discrepancies to future work.

When two techniques are used together, the constraining power increases as expected, with further improvement when all three are used. The ultimate constraining power of the three techniques together demonstrates the limited ability of an EXCLAIM like instrument to put robust constraints on [CII] emission models. However, constraining a larger set of parameters could break degeneracies inherent to the cross-correlation techniques. Stacking, which encapsulates all information in a single number (summed intensity of central voxels),\footnote{Theoretically, one could treat each voxel of the stack as its own observable, like the CVID ratio or cross power spectrum bins are, but we refrain from doing so in this work.} is highly degenerate across the model's parameter space. The CVID and cross power spectrum, however, can separate out the observable information into many bins, making them more sensitive to higher-dimensional parameter spaces. In such a scenario, their abilities to constrain the parameters might differ to such an extent that it becomes more clear which techniques are more useful in extracting information about the signal from LIM maps.

The results of the cross-correlation in the reduced noise scenario of a futuristic EXCLAIM-like scenario largely followed what one would reasonably expect: for 10\% of the noise level, SNR and constraining increased by a factor of about $\sim$10. The present EXCLAIM design serves as a pathfinder for such future experiments and would be capable of achieving the first steps of its astrophysical and cosmological goals. Clearly distinguishing between models is one such goal. In this work, we only considered the \citet[]{padmanabhan_constraining_2019} model, but other models, like the \citet[]{yang_empirical_2022} model also considered by \citet[]{pullen_extragalactic_2023}, are more pessimistic. [CII] emission according to those models would be harder to detect at present noise levels and distinguishing between model parameterizations would pose a significant challenge. Our simulations also neglected broader systematic effects inherent to the data collection process. To make the best use of cross-correlation techniques, robust understanding and cleaning of these effects will be required, and large technological improvements are needed before the reduced noise scenario is achievable. Our results show, however, that there is considerable scientific reward for making those leaps.

Beyond making the case for technological improvements, the reduced noise scenario also makes clear that the CVID ratio cross-correlation technique requires some deeper thought than was initially anticipated before being applied to real data. The high sensitivity to the choice of binning of the intensity map poses a challenge to its consistent application across noise regimes and [CII] emission models. It is unclear at this time what method there can or should be to selecting a binning regime for a LIM map. Requiring different binning regimes also poses a challenge when comparing ratios across signal-to-noise levels, as we see in Figs. \ref{fig:ft_cvids} and \ref{fig:cvid_ratios}. Altering the number of bins and the bin width changes the range and density of Fourier modes that are probed. This, in turn, changes the information that the ratio is able to convey about the underlying distributions. We also note the existence of artifacts in the covariance matrices of the CVID ratio with higher numbers of bins. In the mixed real-imaginary portions, there are "tendrils" of particularly low covariance (see Fig. \ref{fig:ln_cvids}). The origin of these tendrils is unclear, but their presence could be a numerical accident. Regardless, they do make it difficult to invert the covariance matrix to calculate probabilities when constraining $A$.

Ultimately, we might require a different tack altogether to use the CVID. There might be other bases (e.g., wavelet or needlet) beyond Fourier in which to transform the CVID and perform the ratio calculation. Such an alternative might be better at removing theoretically uncorrelated fields and remain well-behaved across signal-to-noise and binning regimes. There might also be some opportunity to use the CVIDs themselves, rather than transforming them to another basis and taking a ratio (or some other similar technique). In lower noise situations, it becomes possible to distinguish the $\ndet{0}$ and $N_{\text{det}}\geq1$ CVIDs (see Fig. \ref{fig:ln_cvids}(b)) themselves, and one could find some metric to measure this difference and extract information.

Implicit in this analysis was the idea that halos containing quasars would obey the same mass-[CII] luminosity relationship as other halos of the same masses but without quasars. This assumption might not necessarily hold, as the energetic processes found in quasars might quench [CII] emission by suppressing star formation rates \citep[]{oei_black_2024, gloudemans_monster_2025}. A reasonable investigation for a future EXCLAIM-like mission would be to differentiate and constrain this difference, such as by using these cross-correlation techniques with a different spectroscopic redshift population to compare the determined galaxy evolution parameters.

Cross-correlating S82 with quasars might require some additional nuance, but there are other opportunities for cross-correlation that could be also be utilized. For high noise instruments like EXCLAIM, one could attempt to cross-correlate with additional datasets. \citet[]{pullen_extragalactic_2023} considered BOSS quasars, and this work uses eBOSS for the one-point statistics. With the Dark Energy Spectroscopic Instrument collaboration releasing their own quasar data sets \citep[]{chaussidon_target_2023}, more objects become available for producing stronger results. Catalogues also exist for other kinds of objects, like HETDEX's Lyman-$\alpha$ emitter catalogue \citep[]{cooper_hetdex_2023}, which overlap S82 and the redshift range of interest. This would eliminate ambiguity regarding how quasars interact with [CII] emission. We leave a robust investigation of alternative surveys for future work, but note that preliminary investigations suggest a HETDEX-like survey would produce similar results to those in this work. Of course, in more futuristic scenarios (with reduced noise or otherwise), one could also target different sky regions or redshifts with their own corresponding catalogues, possibly even in addition to the S82 region.

\section{Conclusion}

In this work, we found that a LIM experiment with the capabilities of EXCLAIM's current design would be capable of detecting [CII] and offer soft constraints on [CII] emission models when its LIM data is cross-correlated with the (e)BOSS QSO survey. The cross-correlation can be done via stacking, the more novel CVID, and the cross power spectrum, which produce SNRs around 4.5, 4, and 8.4, respectively. If one assumes a [CII] emission model like \citet[]{padmanabhan_constraining_2019} for the QSO-containing dark matter halos, these cross-correlation techniques collectively can constrain an amplitude parameter at around 27\% of its fiducial value. Individually, the techniques' constraining powers are around 50\%. These constraints suggest that an EXCLAIM-like instrument would serve as an effective pathfinder for future experiments seeking to measure [CII] emission at these redshifts or using similar technologies.

We also considered a more futuristic scenario in which the noise levels of EXCLAIM's current design were reduced to 10\% of their original. Applying the same techniques, we find that the SNRs increase to about 44, 25, and 34, respectively, with a combined constraining power of around 2\% (about 5\% individually). At that noise level, our capability to understand star formation rates around cosmic noon is drastically increased, highlighting the potential such surveys have for important astrophysical and cosmological work.

Stacking, the CVID, and the cross power spectrum can all be useful tools for analyzing LIM maps. While stacking and the cross power spectrum have been used for LIM previously, we have shown that the CVID can potentially supplement these two techniques, providing even more evidence for detections and offering even greater constraining power on emission models. More work is needed to characterize the properties of the CVID and optimize its application to LIM maps across signal-to-noise levels.

\section*{Acknowledgments}

We would like to thank the entire EXCLAIM team at NASA for their thoughtful discussions and input, and their dedication to LIM science.

SHK was supported by NASA under RFP25\_10-0 issued through the Wisconsin Space Grant Consortium and any opinions, findings, and conclusions or recommendations expressed in this material are those of the authors and do not necessarily reflect the views of the National Aeronautics and Space Administration.

ARP is supported by NASA under award numbers 80NSSC19K1083, 80NSSC22K0666, and 80NSSC25K7478, and by the NSF under award number 2108411. ARP is also supported by the Simons Foundation.

\section*{Data Availability}
The simulated data and simulation code used in this work are not publicly available, but can be made available upon reasonable request to the authors.



\bibliographystyle{mnras}
\bibliography{bib} 




\appendix

\section{On the Limiting Behavior of the CVID Ratio}\label{ap:patrick}
The CVID ratio summary statistic, as described in Eq. \ref{eq:ratio}, is a nonlinear combination of random variables. Thus, even if the uncertainties on the conditional histograms $B_i$ are approximately Gaussian, the statistics of $\mathcal{R}$ may show some divergent behavior. In this appendix, we will use a simplified calculation to attempt to gain some understanding of this nonlinearity, and argue that the estimator remains well-behaved in the regime in which we are using it.

Consider a measurement of $\mathcal{R}(\tilde{I}_i)$ computed at a single Fourier-conjugate-intensity bin $\tilde{I}_i$.  For simplicity let us only consider its real part, as the real and imaginary parts will show qualitatively similar behavior. The measurement will have some expected value $\langle\mathcal{R}\rangle$ and some uncertainty.  For our modeling here, we have assumed that this error is Gaussian, with some standard deviation $\sigma_\mathcal{R}$ that we estimate from simulations. However, from a cursory examination, we can show that this is not formally correct. The denominator of $\mathcal{R}$ is itself a random variable with some expectation value and uncertainty. There will always be a nonzero chance that the uncertainty randomly scatters the denominator arbitrarily close to zero, causing the measured $\mathcal{R}$ to blow up to infinity. This singular behavior can dramatically alter the likelihood.

We will not treat the full problem of ratio distributions here. To gain intuition, we consider a simpler case: a normally-distributed random variable $x$, and a measurement of its inverse $y=1/x$. If $x$ has mean $\mu$ and standard deviation $\sigma$, the expectation value of $y$ is given by
\begin{equation}
\langle y\rangle=\frac{1}{\sqrt{2\pi\sigma^2}}\int_{-\infty}^\infty \frac{1}{x}\exp\left(\frac{(x-\mu)^2}{2\sigma^2}\right)dx.
\end{equation}
This integral of course diverges to infinity at $x=0$. We can obtain a finite value using the Cauchy principal value, which eliminates the singularity:
\begin{equation}
\langle y\rangle = \text{p.v.}\ \frac{1}{\sqrt{2\pi\sigma^2}}\int_{-\infty}^\infty \frac{1}{x}\exp\left(\frac{(x-\mu)^2}{2\sigma^2}\right)dx,
\end{equation}
\begin{multline}
\langle y\rangle=\frac{1}{\sqrt{2\pi\sigma^2}}\lim_{\epsilon\rightarrow0^+}\Biggl[\int_{-\infty}^{-\epsilon} \frac{1}{x}\exp\left(\frac{(x-\mu)^2}{2\sigma^2}\right)dx\\+\int_{\epsilon}^\infty \frac{1}{x}\exp\left(\frac{(x-\mu)^2}{2\sigma^2}\right)dx\Biggr].
\end{multline}
The principal value has closed-form solution
\begin{equation}
\langle y\rangle=\frac{\sqrt{2}}{\sigma}F\left(\frac{\mu}{\sigma\sqrt{2}}\right),
\label{dawson}
\end{equation}
where $F$ is Dawson's function.

\begin{figure}
\centering
\includegraphics[width=\columnwidth]{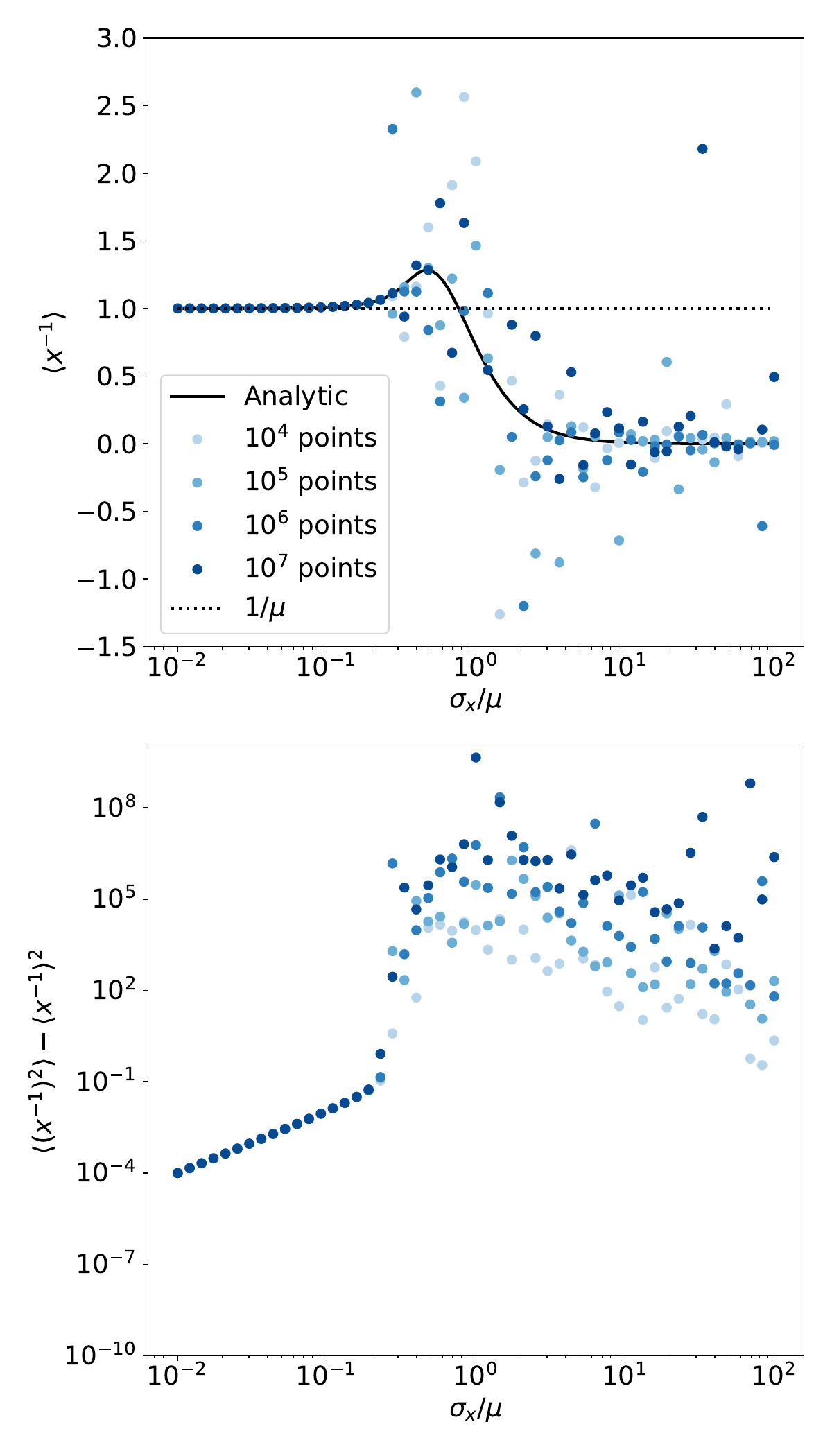}
\caption{(Top) Expectation value of $y=1/x$ where $x$ is a normally distributed random number with mean $\mu=1$ and standard deviation $\sigma$. Computed using the analytic Cauchy principal value from eq. (\ref{dawson}) (black solid line) and simulations with a varying number of random draws (blue points). The value of $1/\langle x\rangle$ is plotted for comparison. (Bottom) Variance of the same set of simulations.}
\label{fig:inverse_demo}
\end{figure}

We can test this with a simple simulation. We draw a large number of normally distributed random numbers with $\mu=1$ at a range of different $\sigma$ values. Then we can directly estimate the mean and variance of $y=1/x$. To test convergence, we run simulations at each $\sigma$ value with $10^4$, $10^5$, $10^6$, and $10^7$ $x$ points. The results are shown in Figure \ref{fig:inverse_demo}. We can see that when $\sigma\ll\mu$, the simulations all converge and $\langle y\rangle=1/\langle x\rangle$.  Once $\sigma$ becomes comparable to $\mu$, however, the expectation value of $y$ starts to change following the Dawson function and the simulations become much noisier.  This can be easily understood following our conceptual sketch from above. When $\sigma\ll\mu$, the odds of any given measurement being scattered close to $x=0$ becomes negligible, and $y$ acts like an approximately Gaussian random variable. When $\sigma\gg\mu$, $x$ is roughly as likely to be positive as negative, so the expectation value drops to zero. It is also quite common to have $x$ close to zero, so the scatter between simulations increases dramatically.

A similar pattern arises if we consider the variance of our simulations. For $\sigma\ll\mu$, there is a well-defined, converged variance which as expected increases with increasing $\sigma$. Once $\sigma$ and $\mu$ become comparable, the simple pattern breaks and we see a large divergence between simulations. Thus we have a perhaps intuitive result: If the signal-to-noise on our denominator is large, the estimate stays well-defined and $\langle y\rangle=1/\langle x\rangle$. Once the signal-to-noise gets too low, the estimator becomes unstable and much harder to use.

The full computation of $\mathcal{R}$ does not exactly match our example. Both the numerator and denominator of $\mathcal{R}$ are random variables, and both are complex at that. However, we can see behavior similar to that displayed in Fig. \ref{fig:inverse_demo} in our simulated $\mathcal{R}$'s. In Fig. \ref{fig:cvid_ratios}, the $\mathcal{R}$ values with and without instrumental noise are in good agreement at low $\tilde{I}$ before eventually diverging. The $\mu/\sigma$ ratio on the histogram in the denominator of $\mathcal{R}$ increases at higher $\tilde{I}$, and is worse in the noise case than the noise-free case. What we are likely seeing then is the transition from Fig. \ref{fig:inverse_demo}. We therefore expect that if we restrict our likelihood calculations to sufficiently low $\tilde{I}$, our $\mathcal{R}$ estimate should stay well-behaved and approximately Gaussian. Convergence tests on the number of $\tilde{I}$ bins included bear this out.

Since we are using simulated estimates for the error on $\mathcal{R}$ here, we consider our empirical approach accurate enough for our forecasts here. We leave a more detailed treatment of $\mathcal{R}$ as a ratio statistic for future work.

\section{Low Noise Figures}\label{ap:lownoise}

In this appendix, we include low-noise versions of figures shown in the main body. Noise has been reduced to 10\% of its original level. 

\begin{figure*}
    \centering
    \includegraphics[width=0.5\textwidth]{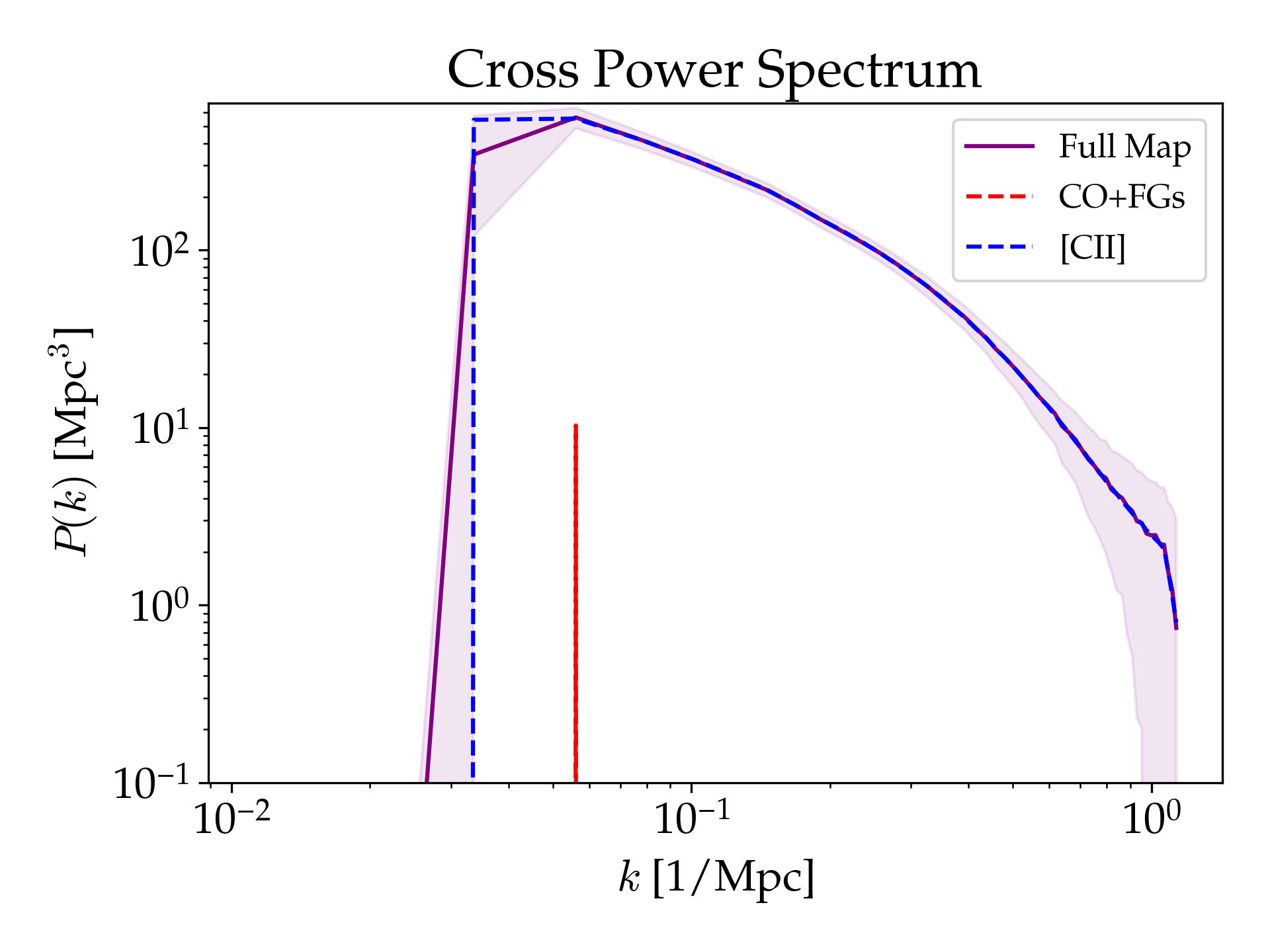}
    \caption{The cross power spectrum between the mock QSO catalogs and simulated LIM maps with reduced noise. The solid purple curve is the cross power for the LIM map with signal and all contaminants included, while the purple dashed curve is for a LIM map with just [CII] signal and the red dashed curve is for a LIM map with just CO interlopers and Galactic foregrounds. The instrument resolution convolves all maps used. The lowest $k_x$, $k_y$, and $k_z$ modes have been filtered from the maps prior to computing the cross power.}
    \label{fig:ln_cross_pow}
\end{figure*}

\begin{figure*}
    \centering
    \begin{subfigure}[t]{0.33\textwidth}
        \centering
        \includegraphics[width=\textwidth]{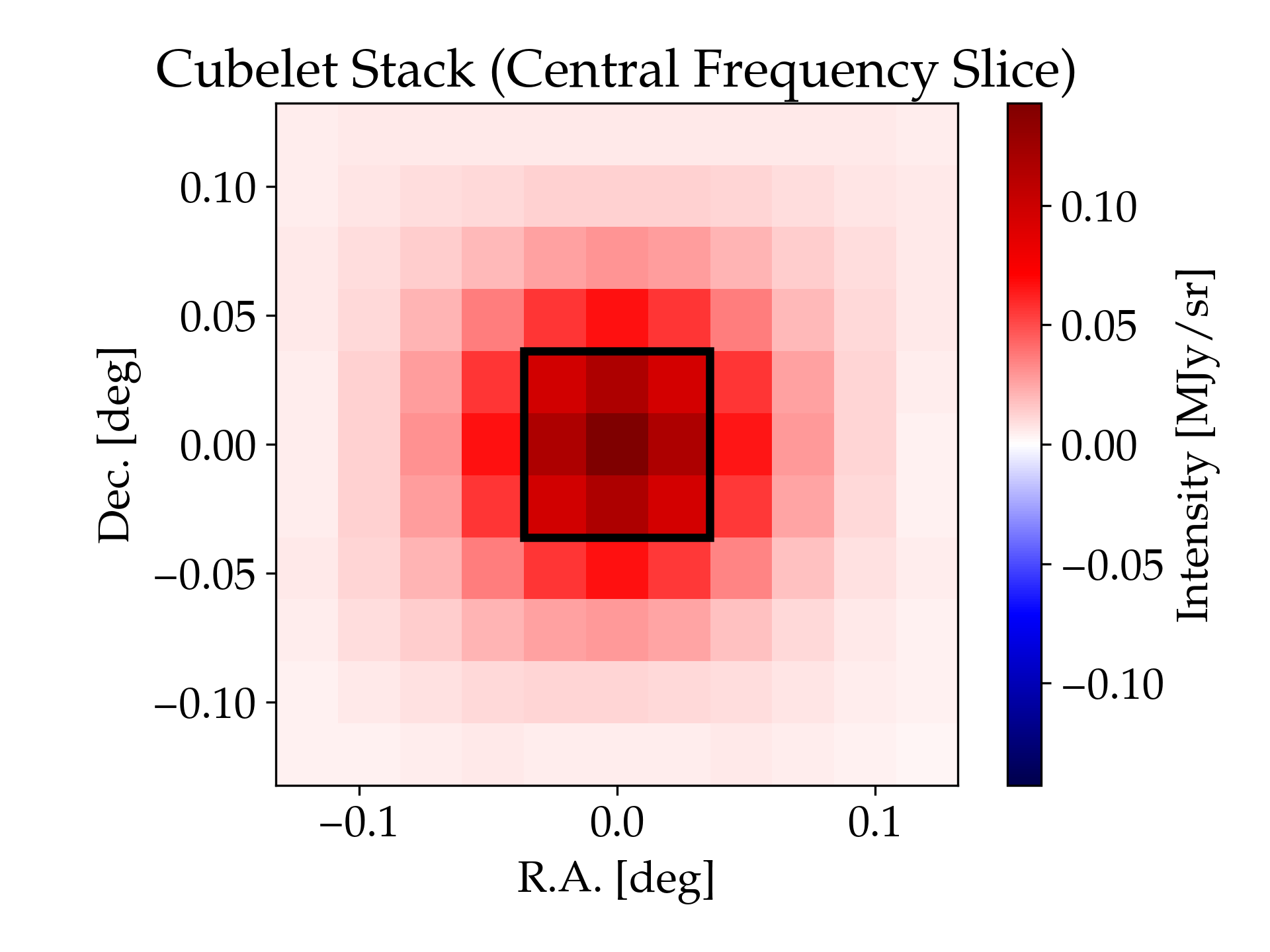}
        \caption{QSO-centered, [CII] only}
    \end{subfigure}%
    ~ 
    \begin{subfigure}[t]{0.33\textwidth}
        \centering
        \includegraphics[width=\textwidth]{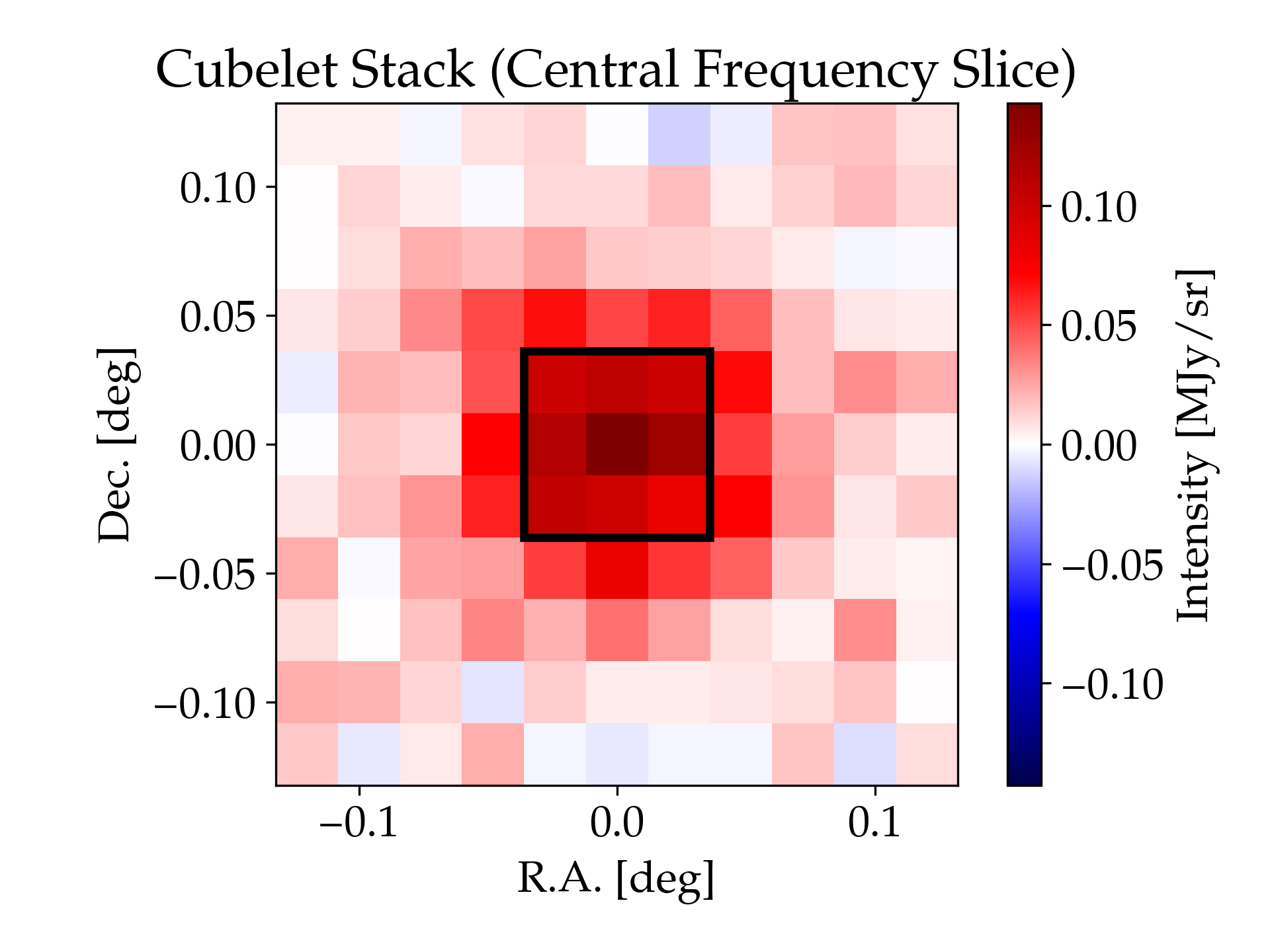}
        \caption{QSO-centered, full map with noise}
    \end{subfigure}%
    ~ 
    \begin{subfigure}[t]{0.33\textwidth}
        \centering
        \includegraphics[width=\textwidth]{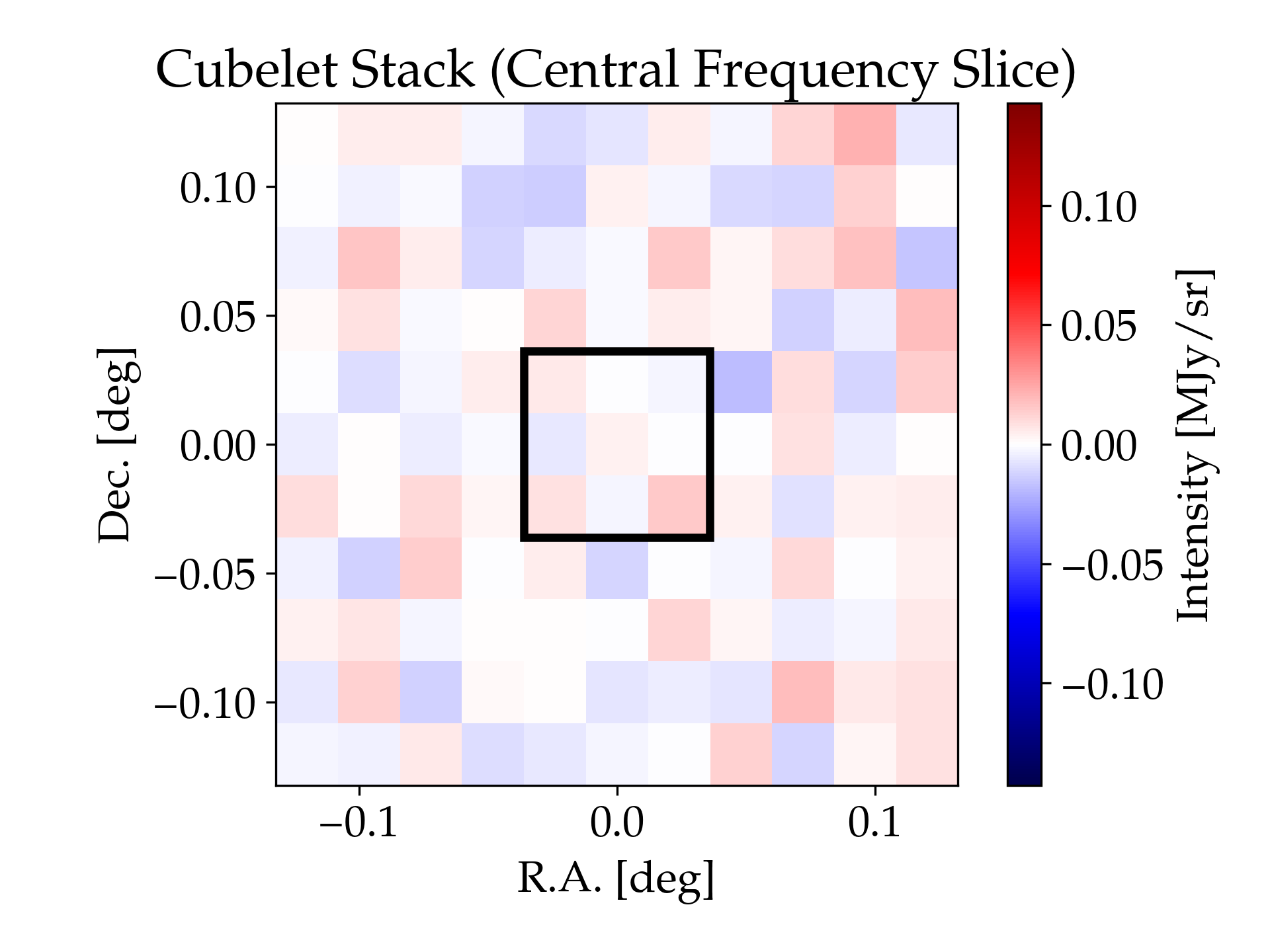}
        \caption{Random locations, full map with noise}
    \end{subfigure}
    \caption{Example stacks resulting from the stacking process for a reduced noise scenario. These depict the frequency slice of the stack in which the QSO (or random location) is located. Coordinates are relative to the center. All stacks are scaled according to the minimum/maximum values of the stack in (b). The black square outlines the central 9 voxels of the stack slice that will be included in the summed intensity observable for the stack.}
    \label{fig:ln_ang_stacks}
\end{figure*}

\begin{figure*}
    \centering
    \begin{subfigure}[t]{0.33\textwidth}
        \centering
        \includegraphics[width=\textwidth]{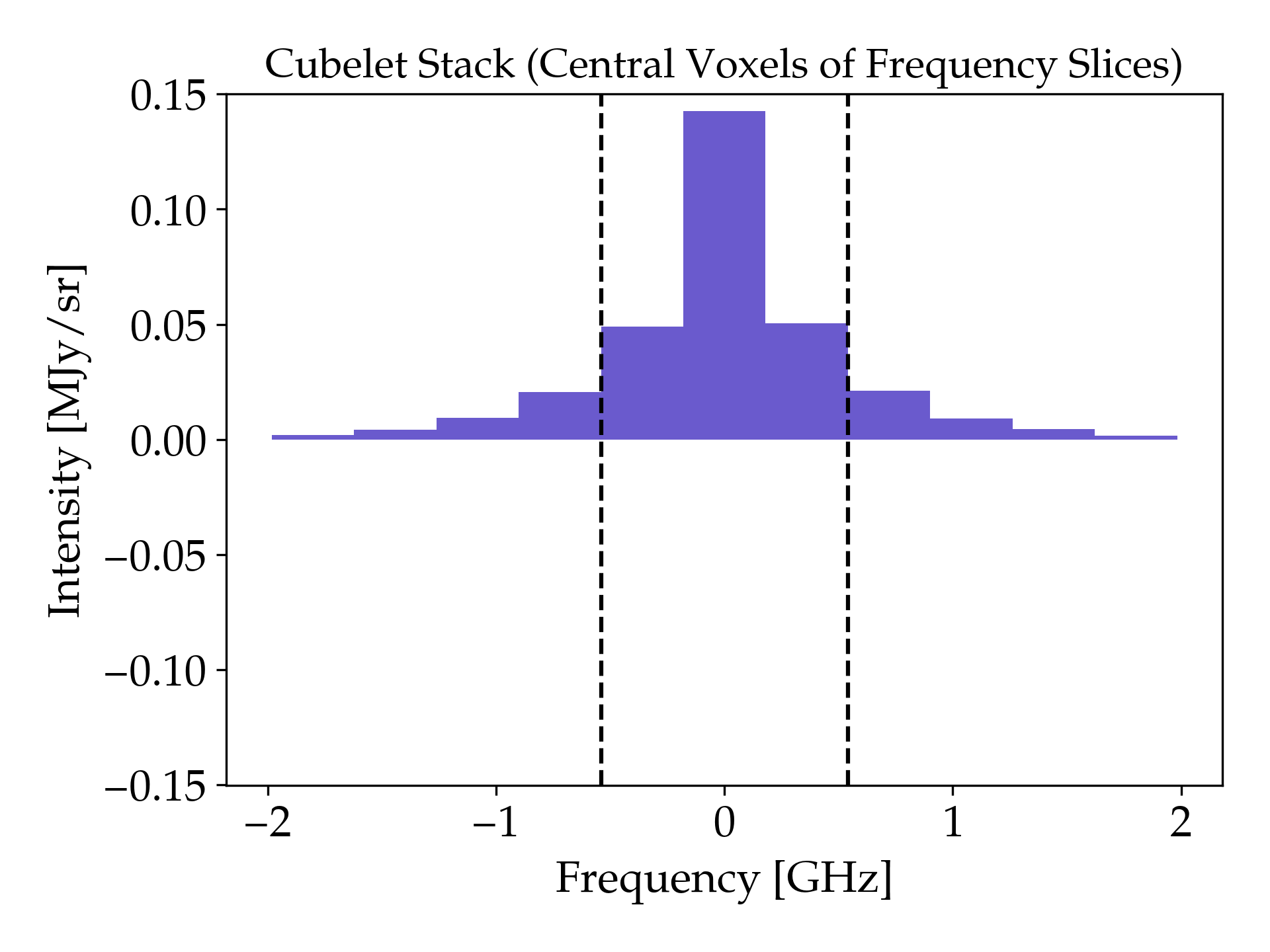}
        \caption{QSO-centered, [CII] only}
    \end{subfigure}%
    ~ 
    \begin{subfigure}[t]{0.33\textwidth}
        \centering
        \includegraphics[width=\textwidth]{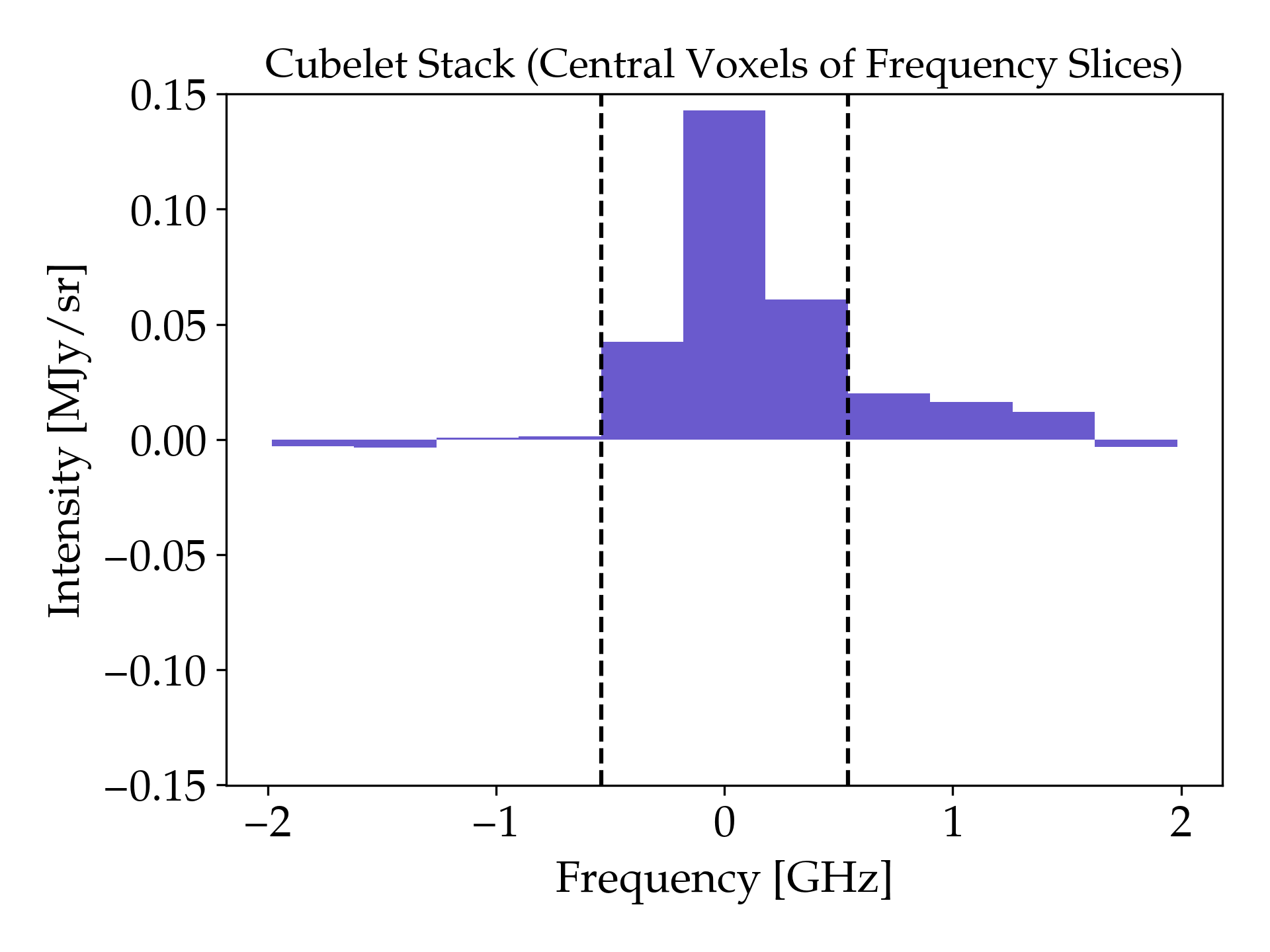}
        \caption{QSO-centered, full map with noise}
    \end{subfigure}%
    ~ 
    \begin{subfigure}[t]{0.33\textwidth}
        \centering
        \includegraphics[width=\textwidth]{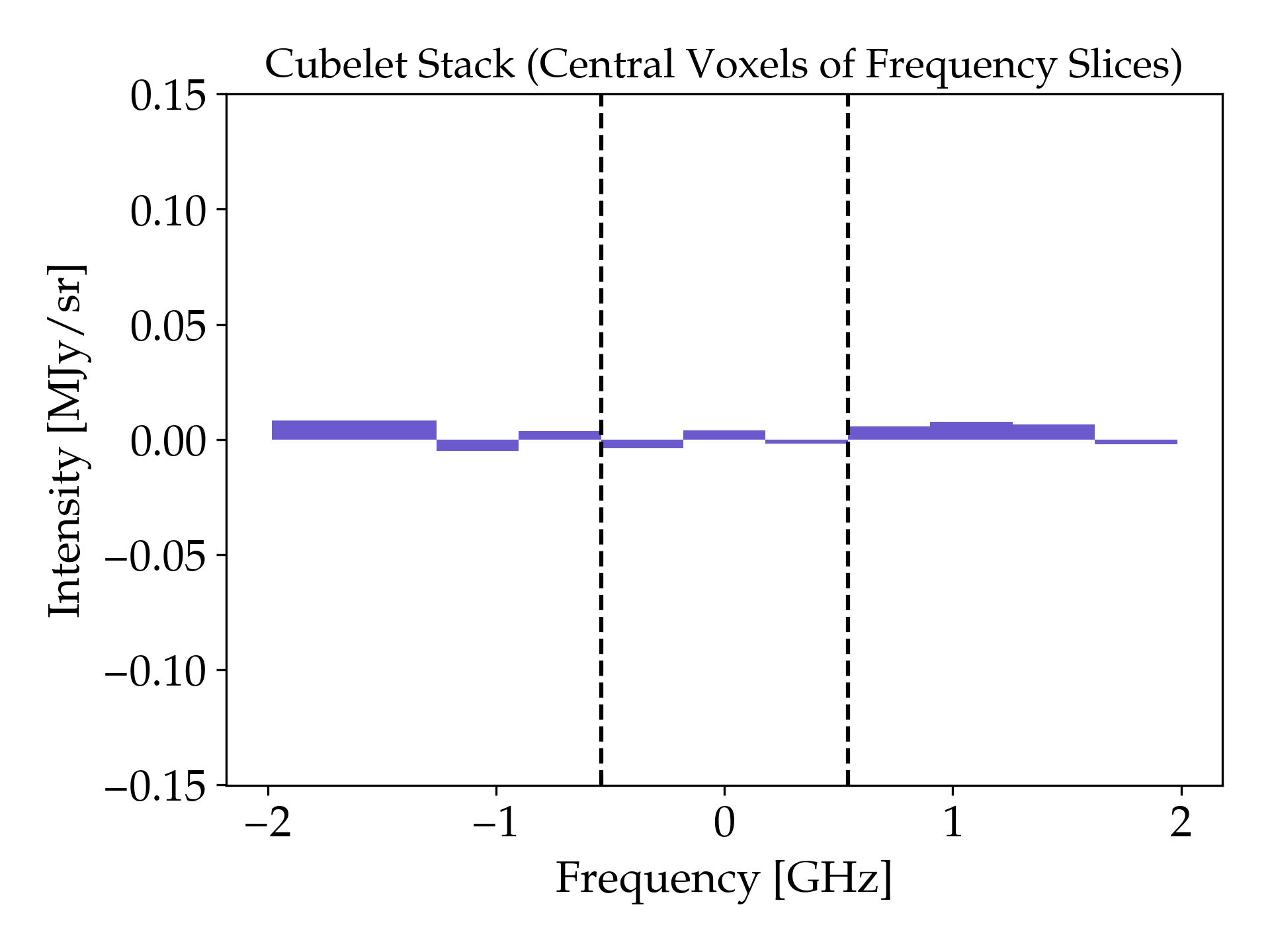}
        \caption{Random locations, full map with noise}
    \end{subfigure}
    \caption{Example stacks resulting from the stacking process for a reduced noise scenario. These depict the central voxels of the stack across frequency slices. Coordinates are relative to the center. All stacks are scaled according to the minimum/maximum values of the stack in (b). The black dashed lines outline the central 3 voxels of the stack  that will be included in the summed intensity observable for the stack.}
    \label{fig:ln_freq_stacks}
\end{figure*}

\begin{figure*}
    \centering
    \begin{subfigure}[t]{0.49\textwidth}
        \centering
        \includegraphics[width=\textwidth]{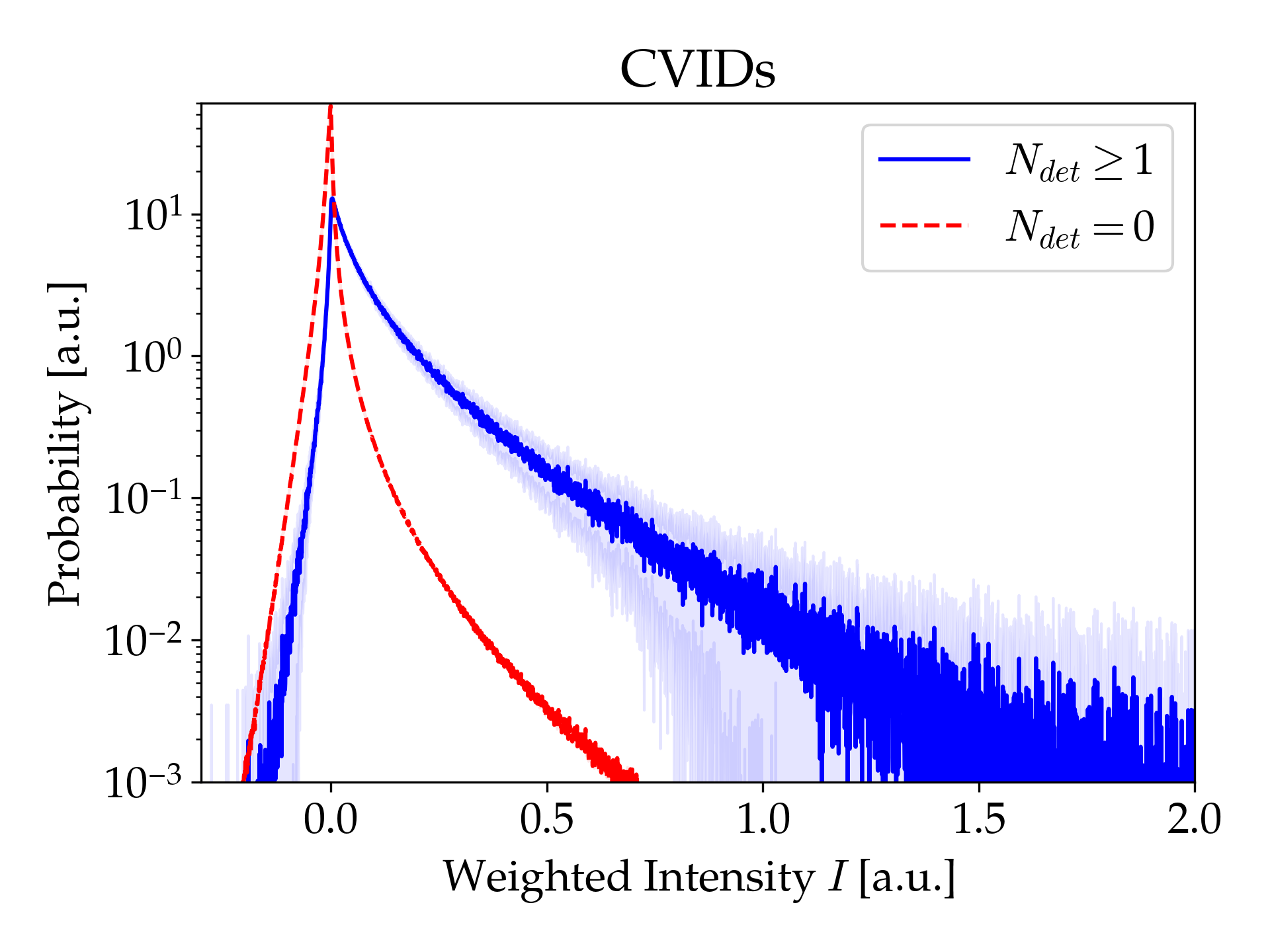}
        \caption{[CII] Only}
    \end{subfigure}%
    ~ 
    \begin{subfigure}[t]{0.49\textwidth}
        \centering
        \includegraphics[width=\textwidth]{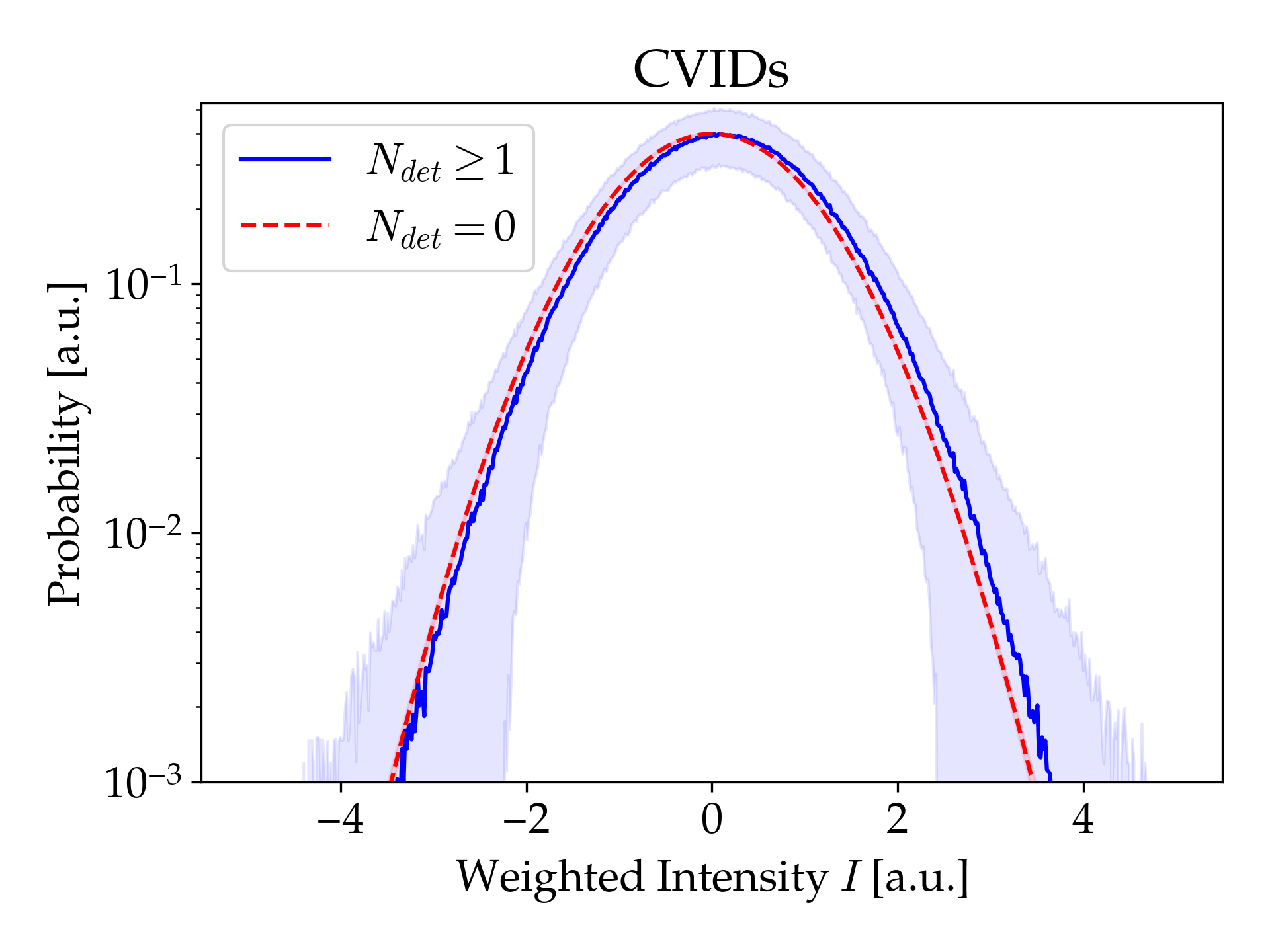}
        \caption{Full map with noise}
    \end{subfigure}
    \caption{The mean CVIDs produced from the reduced noise LIM maps. The solid blue curve is the mean CVID for voxels containing a QSO and the immediately adjacent voxels. The dashed red curve is the mean CVID for all other voxels. Each curve is accompanied by a shaded region of the same color showing the one standard deviation error of each bin. Note that the $N_{\text{det}}=0$ curves' errors are too small to be seen. In (b), bins are omitted at regular intervals for ease of visualization. This omission is not utilized in any calculations.}
    \label{fig:ln_cvids}
\end{figure*}


\begin{figure*}
    \centering
    \begin{subfigure}[t]{0.49\textwidth}
        \centering
        \includegraphics[width=\textwidth]{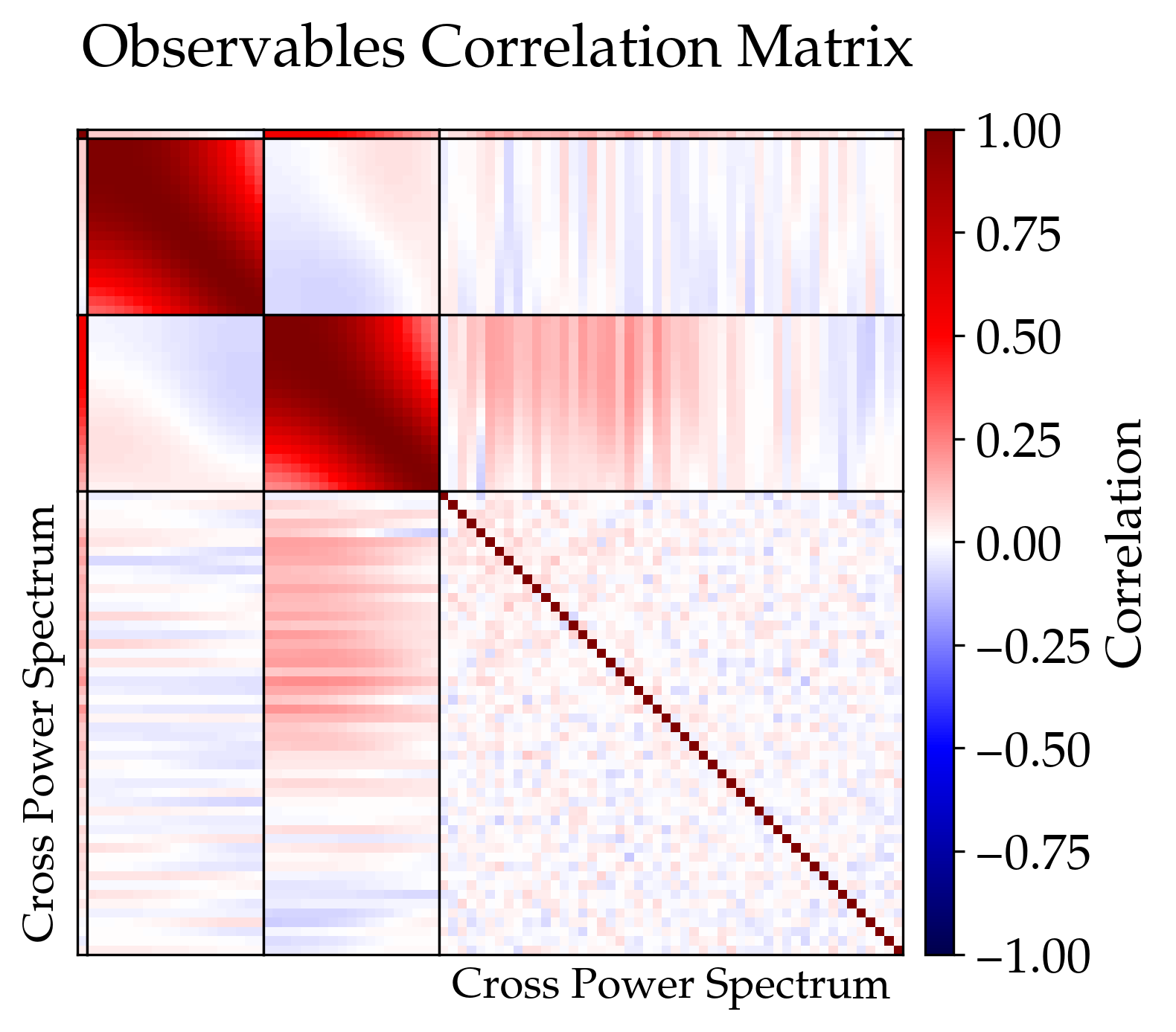}
        \caption{Full Correlation}
    \end{subfigure}%
    ~ 
    \begin{subfigure}[t]{0.49\textwidth}
        \centering
        \includegraphics[width=\textwidth]{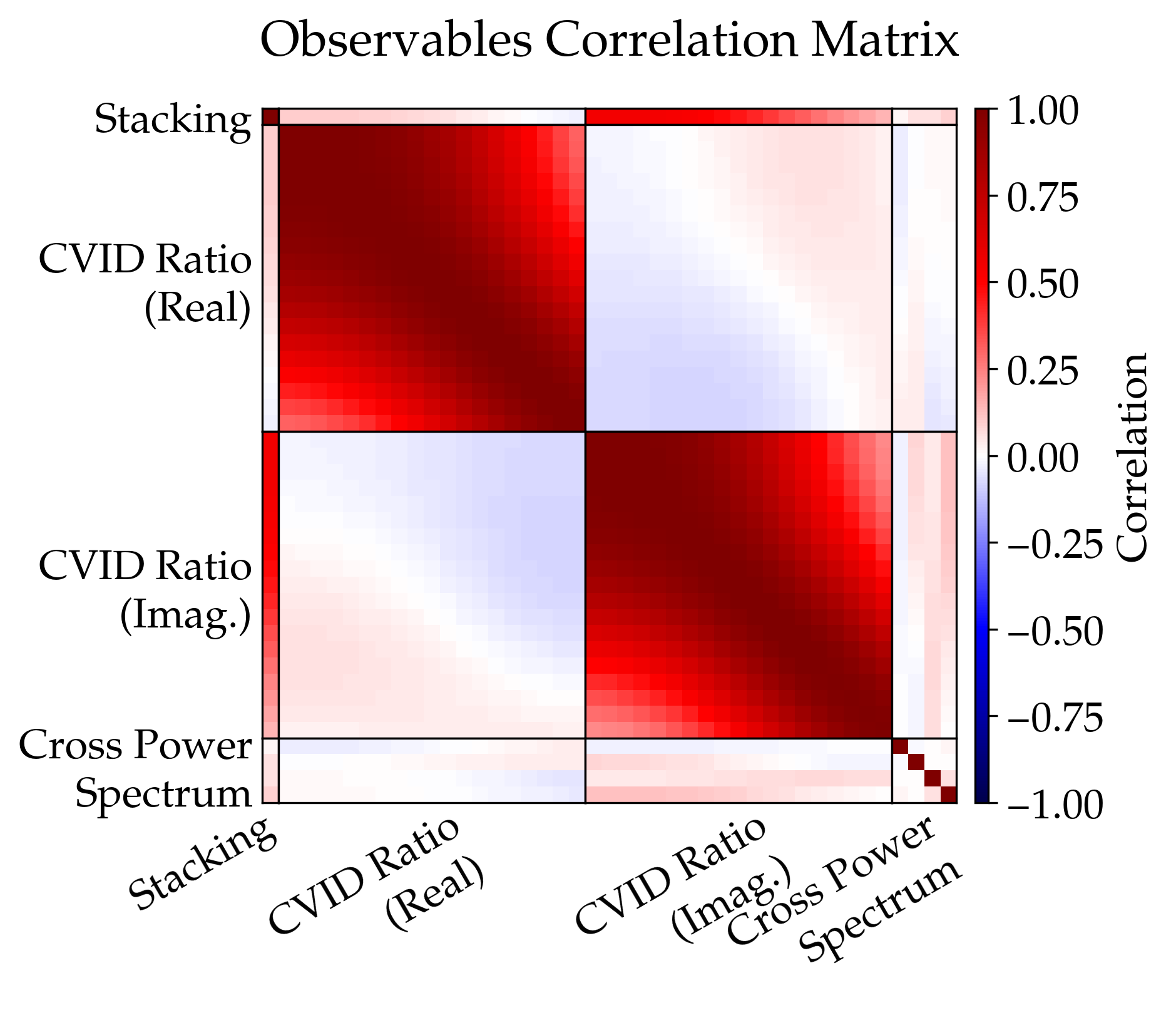}
        \caption{Upper Left Corner}
    \end{subfigure}
    \caption{The correlation matrix of the three observables as described in Sec. \ref{sec:disc} for the reduced noise LIM map. Each pixel represents one bin of the observable. (a) is the full correlation matrix, while (b) is the upper left corner of (a) enlarged for visibility. The correlation matrix is unitless and can range from $+1$ (directly correlated) to $-1$ (inversely correlated).}
    \label{fig:ln_corr}
\end{figure*}


\bsp	
\label{lastpage}
\end{document}